\RequirePackage{fix-cm}
\documentclass[twocolumn, epjc3, comma, sort&compress, natbib]{svjour3}
\bibpunct{[}{]}{,}{n}{}{,} 

\journalname{Eur. Phys. J. C}

\usepackage{latexsym}
\usepackage{amsmath}
\usepackage{amssymb}
\usepackage{amsfonts}
\usepackage{comment}
\usepackage{CJKutf8}

\usepackage[mathscr,scaled=1.15]{urwchancal}
\DeclareFontFamily{OT1}{pzc}{}
\DeclareFontShape{OT1}{pzc}{m}{it}%
{<-> s * [1.15] pzcmi7t}{}
\DeclareMathAlphabet{\mathpzc}{OT1}{pzc}{m}{it}

\usepackage{color}

\usepackage{supertabular}
\usepackage{placeins}
\usepackage{epsfig}
\usepackage{graphicx}

\definecolor{purple}{rgb}{0.5,0,0.5}
\definecolor{blue}{rgb}{0.0,0,0.9}
\definecolor{prdblue}{rgb}{0.133,0.118,0.498}
\usepackage[colorlinks=true, pdfstartview=FitV, linkcolor=prdblue, citecolor= prdblue, urlcolor=prdblue]{hyperref}

\begin{document}
\begin{CJK*}{UTF8}{gbsn}

\title{$\,$\\[-6ex]\hspace*{\fill}{\normalsize{\sf\emph{Preprint no}.\
NJU-INP 123/26}}\\[1ex]
Neutral Pseudoscalar Mesons, Hadron Light-by-Light Scattering, and the Muon Anomalous Magnetic Moment}

\author{Cong-Cong Ye (叶聪聪)\thanksref{NJU,INP}%
       $\,^{\href{https://orcid.org/0009-0004-4621-1732}{\textcolor[rgb]{0.00,1.00,0.00}{\sf ID}}}$
\and
        Yu-Chen Hu（胡宇琛)\thanksref{NJU,INP}%
       $\,^{\href{https://orcid.org/0009-0004-8330-1454}{\textcolor[rgb]{0.00,1.00,0.00}{\sf ID}}}$
\and
    \\Minghui Ding (丁明慧)\thanksref{NJU,INP}%
       $\,^{\href{https://orcid.org/0000-0002-3690-1690}{\textcolor[rgb]{0.00,1.00,0.00}{\sf ID}}}$
\and 
    Craig D. Roberts\thanksref{NJU,INP}%
       $\,^{\href{https://orcid.org/0000-0002-2937-1361}{\textcolor[rgb]{0.00,1.00,0.00}{\sf ID}}}$
}

\authorrunning{Cong-Cong Ye \emph{et al}.} 

\institute{School of Physics, Nanjing University, Nanjing, Jiangsu 210093, China \label{NJU}
           \and
           Institute for Nonperturbative Physics, Nanjing University, Nanjing, Jiangsu 210093, China \label{INP}
\\[1ex]
Email:
\href{mailto:mhding@nju.edu.cn}{mhding@nju.edu.cn} (MD);
\href{mailto:cdroberts@nju.edu.cn}{cdroberts@nju.edu.cn} (CDR)
            }

\date{2026 December 01}  

\maketitle
\end{CJK*}

\begin{abstract}
%
Continuum Schwinger function methods\linebreak (CSM) are used to provide estimates of pseudoscalar meson pole contributions -- $\mathsf P = \pi^0$, $\eta$, $\eta^\prime$, $\eta_c$, $\eta_b$ -- to the muon anomalous magnetic moment.  
The analysis framework is benchmarked via successful applications to an array of other meson and nucleon properties.
It uses the leading order CSM truncation, augmented by a parametrisation of non-Abelian anomaly effects, to deliver an internally consistent treatment of all the pseudoscalar--two-photon transition form factors required to complete such a calculation.  
Employing those form factor predictions, the following results are obtained (units $10^{-11}$):
$a_\mu^{\pi^0} +
a_\mu^{\eta}  +
a_\mu^{\eta^\prime} = 84.3(0.8)$ -- a value lower than some other contemporary estimates;
$a_\mu^{\eta_c} = 0.82(0.04)$;
and a value of $a_\mu^{\eta_b}$ that is negligible.
The total contribution is $\sum _{\mathsf P} a_\mu^{\mathsf P} = 85.1(0.8) / 10^{11}$.
Our analysis suggests that this pole contribution to $a_\mu$ is still not known with precision better than 5\%.
\end{abstract}


\section{Introduction}
\label{secintro}
Predictions made by the Standard Model of Particle Physics (SM) provide the benchmark against which sear\-ches for ``new physics'' are judged; so measurements of the $\mu$-lepton anomalous magnetic moment, $a_\mu = (g-2)_\mu/2$, which lay above SM predictions by $2.2$-$2.7\,\sigma$ \cite{Muong-2:2006rrc}, generated a great deal of interest \cite{Aliberti:2025beg}.
The situation has calmed somewhat following the more recent measurement \cite{Muong-2:2025xyk}, which reported
\begin{equation}
a_\mu = 0.001165920705(114)\,,
\end{equation}
and updates to the SM estimate, summarised, \emph{e.g}., in Ref.\,\cite[Table~2]{Hertzog:2025ssc}: now, experiment and theory agree within $1\,\sigma$.

\begin{figure}[b]
\centerline{%
\includegraphics[clip, width=0.475\textwidth]{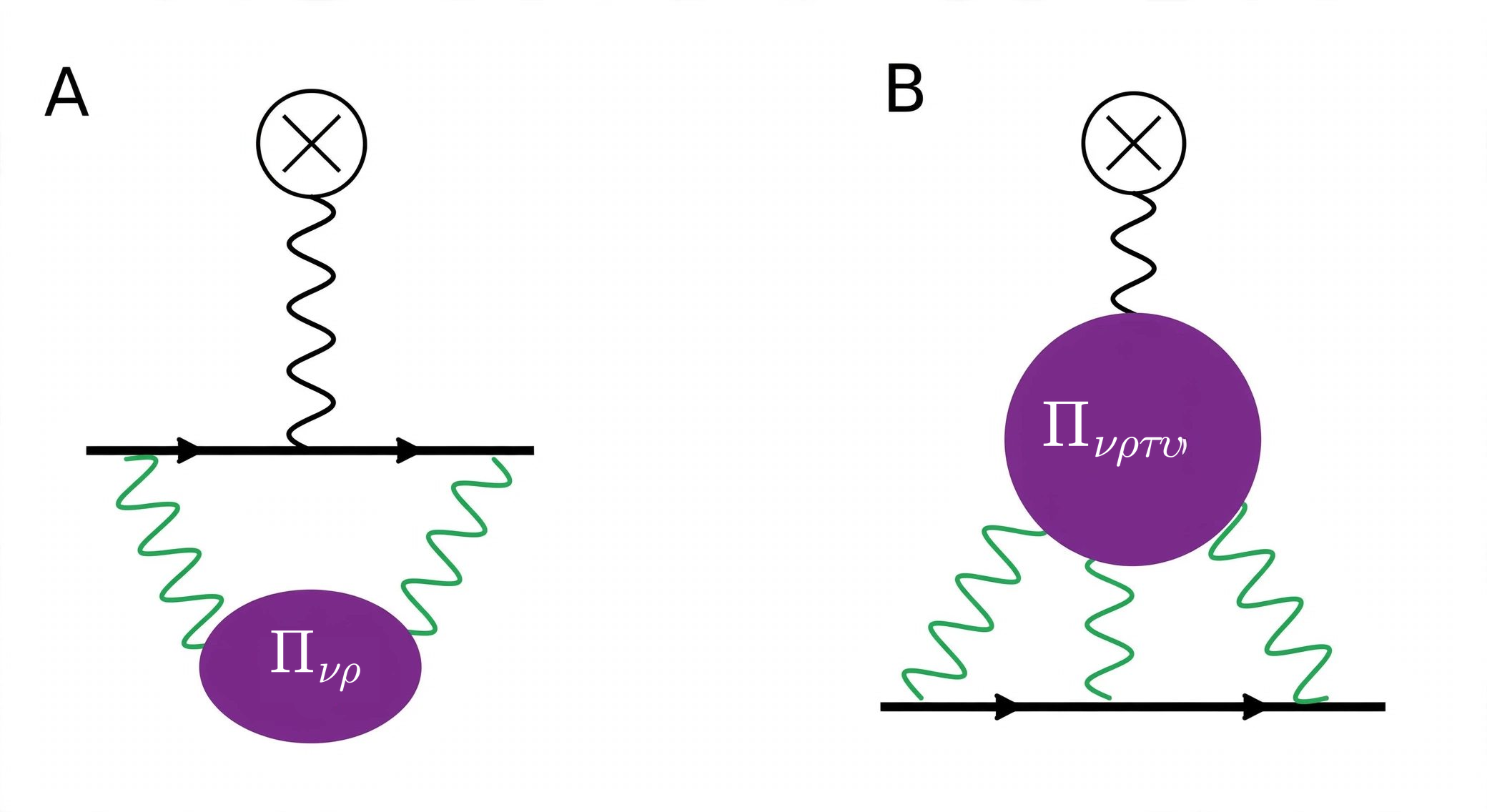}}
\vspace*{-2ex}

\caption{\label{F1}
Hadron vacuum polarisation ({\sf A}) and light-by-light scattering ({\sf B}) contributions to the $\mu$-lepton anomalous magnetic moment, both of which must be estimated using nonperturbative methods.
Legend.
Undulating lines -- photons, real (black) when external and virtual (green) when internal.
Straight lines -- $\mu$-leptons.
Filled (purple) vertices -- hadron interaction pieces, the dominant parts of which must be identified and calculated.
}
\end{figure}

Notwithstanding, there are many parts to the SM theory estimate \cite[Table~2]{Hertzog:2025ssc}.
Principal amongst the pieces that require nonperturbative calculation are the hadron vacuum polarisation (HVP) contribution and the hadron light-by-light (HLbL) scattering component, which are illustrated in Fig.\,\ref{F1}.
Today, the estimated uncertainty on the HVP result is less than 1\% \cite{Boccaletti:2024guq}, whereas that on HLbL is $8.5$\% \cite{Hertzog:2025ssc}.
Given these circumstances, herein we focus on the HLbL contribution.

\begin{figure}[b]
\hspace*{-1.ex}{%
\includegraphics[clip, width=0.525\textwidth]{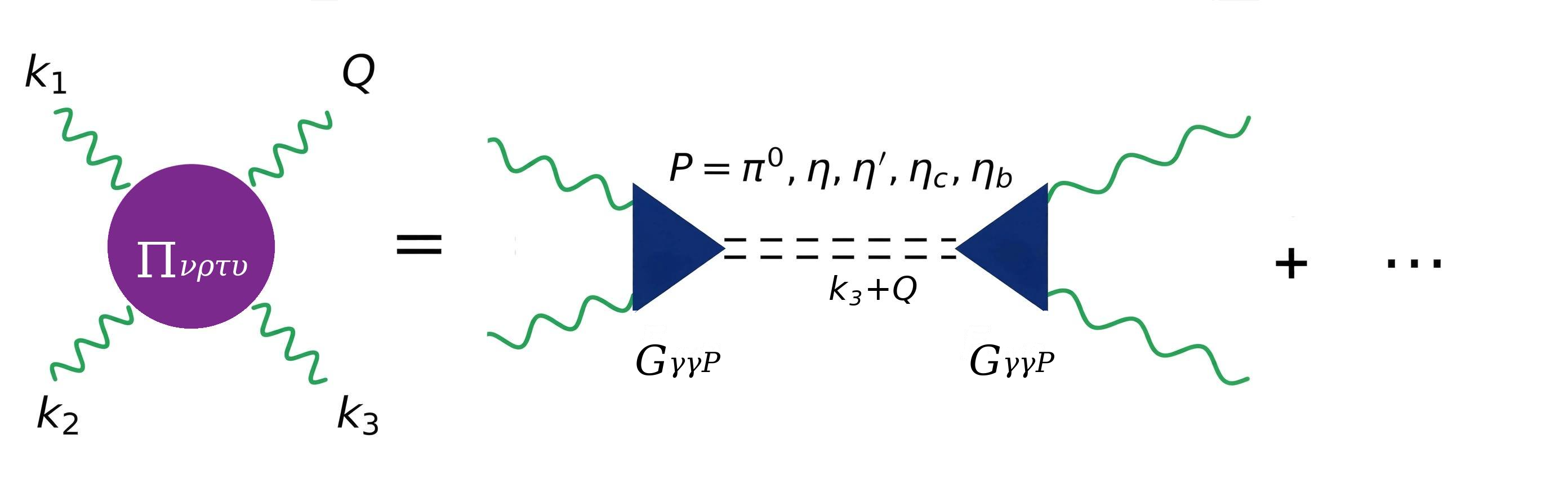}}
\vspace*{-2ex}

\caption{\label{F2tag}
Pseudoscalar meson pole contributions dominate in hadron light-by-light scattering: $k_1+k_2+k_3+Q=0$.
Legend.
Dark-blue vertices -- (anomalous) pseudoscalar to two-photon structure/form factors;
dashed double line -- intermediate-state pseudoscalar meson propagator;
 ``$\ldots$" -- other two permutations of $k_{1,2,3}$.
}
\end{figure}

Estimates of the HLbL contribution have a long history.
The dominant piece is the pseudoscalar meson pole term \cite[Sec.\,4.2.4]{Aoyama:2020ynm}: $\gamma\gamma \to {\mathsf P} \to \gamma\gamma$, where the intermediate state ${\mathsf P} = \pi^0, \eta, \eta^\prime, \eta_c, \eta_b$; see Fig.\,\ref{F2tag}.
In most analyses, only $\pi^0, \eta, \eta^\prime$ are considered; in the present context, Refs.\,\cite{Goecke:2010if, Goecke:2012qm} are pertinent and pioneering.
However, Ref.\,\cite{Raya:2019dnh}, also using continuum Schwinger function methods (CSMs) \cite{Roberts:2021nhw, Ding:2022ows, Raya:2024ejx}, included, for the first time, both $\eta_c$ and $\eta_b$ intermediate states.
It found that the $\eta_c$ contribution is of the same order as the uncertainty typically associated with the net $\pi^0, \eta, \eta^\prime$ contribution.  Thus, Ref.\,\cite{Raya:2019dnh} suggests that this contribution should not be neglected.

As the first (and hitherto only) study to include $\eta_c, \eta_b$, a number of simplifications were made in Ref.\,\cite{Raya:2019dnh}.
For instance, 
perturbation theory integral representations (PTIRs) \cite{Nakanishi:1969ph, Nakanishi:1971, Chang:2013nia} were used for all quark propagators and pseudoscalar meson bound-state amplitudes that are needed to compute the ${\sf P}\gamma^{(\ast)}\gamma^{(\ast)}$ form factors, $F_{{\sf P}\gamma\gamma}$.
Furthermore, ``gauge-technique'' \emph{Ans\"atze} \cite{Delbourgo:1977jc, Ding:2018xwy} were used to parametrise the photon-quark vertices that are also required in the calculation of $F_{{\sf P}\gamma\gamma}$, with parameters tuned to reproduce the $\pi^0$ anomaly constraint at zero momentum transfer \cite{Adler:1969gk, Bell:1969ts, Roberts:1994hh, Holl:2005vu} and empirical two-photon decay widths for the other pseudoscalar mesons.

In revisiting the Ref.\,\cite{Raya:2019dnh} analysis herein, those expedients are eschewed.
Instead, we calculate all propagators, amplitudes, and vertices directly,
using the leading-order symmetry-preserving CSM truncation scheme\linebreak
(rainbow-ladder truncation \cite{Munczek:1994zz, Bender:1996bb}, RL) and, where necessary, the Schlessinger point method (SPM) -- see, \emph{e.g}.,  Refs.\,\cite{PhysRev.167.1411, Schlessinger:1966zz, Tripolt:2016cya, Yao:2024ixu, Cheng:2024cxk} -- for extrapolation of transition form factor results onto otherwise inaccessible domains required for the estimation of $\sf P$-meson contributions to $a_\mu$.  
Our formulation and solution of the problem is fully benchmarked by successful predictions for an array of pseudoscalar meson and nucleon properties; see, e.g., Refs.\,\cite{Chen:2018rwz, Ding:2018xwy, Qin:2018dqp, Yao:2021pyf}.

At this point it is worth mentioning that, after thirty years of use, RL truncation is known to be reliable for pseudoscalar meson ground-states.  (Owing to the non-Abelian anomaly \cite{Christos:1984tu}, the $\eta$-$\eta^\prime$ complex is a special case.  It requires a correction to RL truncation, which is understood \cite{Ding:2018xwy} and used below.)
This has been established practically, through widespread applications \cite{Roberts:2021nhw, Ding:2022ows, Raya:2024ejx}, and via the use of higher-order truncations, which has revealed that the cumulative effect of improvement to RL truncation in these channels can be absorbed into a modest modification of the quark + quark scattering kernel \cite{Chang:2013pq, Binosi:2014aea, Williams:2015cvx, Binosi:2016rxz, Binosi:2016wcx, Xu:2022kng}.
%


The discussion is arranged as follows.
Section~\ref{sectwo} introduces the basic formulae that are used to estimate pseudoscalar meson pole HLbL contributions to the muon anomalous magnetic moment and their relationship to the (anomalous) ${\mathsf P}\gamma^{(\ast)}\gamma^{(\ast)}$ transition form factors.
Sections~\ref{SecThree}\,--\,\ref{Secetab} describe, in turn, our approach to calculating $F_{{\sf P}\gamma\gamma}$, ${\mathsf P}=\pi^0, \eta, \eta^\prime, \eta_c, \eta_b$.  
The $\pi^0$ is the archetypal case, Sec.\,\ref{SecThree}; but in analysing some of the other states, a few issues must be resolved, e.g., 
the non-Abelian anomaly complicates the treatment of $\eta$, $\eta^\prime$, Sec.\,\ref{etaprimeeta},
and the fact that the charm quark is neither light nor truly heavy introduces challenges for the $\eta_c$, Sec.\,\ref{Secetac}.
The calculation is more straightforward for the $\eta_b$, Sec.\,\ref{Secetab}, which is constituted from heavy valence degrees of freedom.  
Section~\ref{epilogue} provides a summary and perspective.

\section{Pseudoscalar Meson Contribution  to Light-by-Light scattering}
\label{sectwo}
In any calculation of the muon anomalous magnetic moment, one is interested in the $\gamma\mu$ interaction vertex:
\begin{equation}
\Gamma_\upsilon^{\gamma\mu}(p,p^\prime) = i e
\Lambda_+(p) \bigg[
F_1(t) \gamma_\upsilon  + \frac{Q_\tau}{2m_\mu} \sigma_{\upsilon\tau} F_2(t) \bigg]
\Lambda_+(p^\prime)\,,
\label{mucurrent}
\end{equation}
where  $e$ ($m_\mu$) is the muon electric charge (mass);\linebreak
%
$\Lambda_+(p^{(\prime)}) = (m_\mu - i \gamma\cdot p^{(\prime)})/(2 m_\mu)$ are positive energy muon projection operators;
$F_{1,2}$ are, respectively, the $\mu$-lepton Dirac and Pauli form factors;
and $Q=p^\prime - p$ is the momentum transfer, with $t=Q^2$.
In this case, one has the following kinematic (elastic scattering) constraints:
\begin{equation}
p^2 = -m_\mu^2,
\quad
2 p\cdot Q + Q^2 = 0 .
\label{kinematics}
\end{equation}
(We use a Euclidean-space formulation of the problem \cite[Sec.\,1]{Ding:2022ows}, with the metric conventions defined, e.g., in Ref.\,\cite[Appendix\,B]{Segovia:2014aza}.)
Of course, charge conservation entails $F_1(0)=1$; and the anomalous magnetic moment is
\begin{equation}
a_\mu = F_2(t=0)\,.
\end{equation}
Since one can approach this limit from spacelike momentum transfers, it is sufficient to consider $Q=0$ in Fig.\,\ref{F2tag}.

Consider the HLbL contribution to $a_\mu$ involving\linebreak $\Pi_{\nu\rho\sigma\tau}(k_1,k_2,k_3,Q)$  in Fig.\,\ref{F1}B.  Whereas the total contribution is finite, the mutiloop structure entails that individual contributions exhibit ultraviolet $\ln$-divergences \cite{Aldins:1970id}.
This complication can be circumvented by exploiting the following (current-conservation) Ward-Green-Takahashi identity:
\begin{equation}
Q_\upsilon \Pi_{\nu\rho\tau\upsilon}(k_1,k_2,k_3,Q) = 0\,,
\end{equation}
for then it is straightforward to show that
\begin{subequations}
\label{rebuild}
\begin{align}
\Pi_{\nu\rho\tau\upsilon}(k_1,k_2,k_3,Q) &
= - Q_\kappa \Pi_{(\upsilon)\nu\rho\tau\kappa}(k_1,k_2,k_3,Q) , \\
\Pi_{(\upsilon)\nu\rho\tau\kappa}(k_1,k_2,k_3,&Q) =
\frac{\partial}{\partial Q_\upsilon} \Pi_{\nu\rho\tau\kappa}(k_1,k_2,k_3,Q) .
\end{align}
\end{subequations}
It is furthermore worth highlighting that \cite{Aldins:1970id, Colangelo:2015ama}:
\begin{equation}
\Pi_{\nu\rho\tau\upsilon}(k_1,k_2,k_3,Q) \stackrel{Q \simeq 0}{\propto} |Q| ;
\label{PiZero}
\end{equation}
namely, $\Pi_{\nu\rho\tau\upsilon}(k_1,k_2,k_3,Q) $ vanishes linearly with $Q_{\upsilon}$, so the Eq.\,\eqref{rebuild} derivative is regular in the neighbourhood $|Q|\simeq 0$.
(Similar statements pertain to the other argument vectors, owing to Bose symmetry.)

Subsequently defining all relevant integrals via\linebreak $\Pi_{(\upsilon)\nu\rho\tau\kappa}(k_1,k_2,k_3,Q)$, then each individual term is finite and so is its contribution to $a_\mu$ when reconstructed via Eq.\,\eqref{rebuild}.
Namely,
\begin{equation}
\Gamma_\upsilon^{\gamma\mu}(p,p^\prime) = - Q_\kappa \Gamma_{(\upsilon)\kappa}^{\gamma\mu}(p,p^\prime) ,
\label{vertex}
\end{equation}
with
\begin{align}
\Gamma_{(\upsilon)\kappa}^{\gamma\mu}(p,p^\prime)
& = i e^3  \int_{dk_1}\int_{dk_2}
\Pi_{(\upsilon) \nu\rho\tau\kappa}(k_1,k_2,k_3,Q) \nonumber \\
&
\quad \times {\mathsf D}_{\nu\nu^\prime}(k_1) {\mathsf D}_{\rho\rho^\prime}(k_2) {\mathsf D}_{\tau\tau^\prime}(k_3) \nonumber \\
&
\quad \times \Lambda_+(p) \gamma_{\nu^\prime} {\mathsf S}(p_1) \gamma_{\rho^\prime} {\mathsf S}(p_2) \gamma_{\tau^\prime} \Lambda_+(p^\prime),
\label{Picoupling}
\end{align}
where
$\int_{dk}=\int d^4k/(2\pi)^4$ is a four-dimensional momentum integral;
${\mathsf D}_{\nu\nu^\prime}$, etc., are photon $2$-point functions (propagators);
${\mathsf S}(p_{1,2})$ are $\mu$-lepton propagators; and
\begin{subequations}
\label{momenta}
\begin{align}
p_1 & = p - k_1 , \\
p_2 & = p - k_1 -k_2 , \\
k_3 & = -k_1-k_2-Q .
\end{align}
\end{subequations}

Returning now to Eq.\,\eqref{mucurrent}, define the following (projection) matrices:
\begin{subequations}
\label{projectors}
\begin{align}
2 i e {\mathbb P}_\upsilon^1 & = \gamma_\upsilon , \\
2 i e {\mathbb P}_\upsilon^2 &= 3 i e  {\mathbb P}_\upsilon^1 +
\frac{m_\mu Q_\kappa}{Q^2} \sigma_{\upsilon\kappa} .
\end{align}
\end{subequations}
With this choice and using Eq.\,\eqref{kinematics}, one has (with the trace over spinor indices):
\begin{equation}
F_2(0) = \left. {\rm tr}_{\rm D} {\mathbb P}_\upsilon^2 \Gamma_\upsilon^{\gamma\mu}(p,p^\prime)
\right|_{t=0} .
\end{equation}

Following Ref.\,\cite{Raya:2019dnh} (and numerous earlier studies), we approximate Eq.\,\eqref{Picoupling} by the pseudoscalar meson pole contributions depicted in Fig.\,\ref{F2tag}:
\begin{align}
& \Pi_{\nu\rho\tau\upsilon}  =
{\cal G}_{\nu\rho}^{\mathsf P}(k_1,k_2) \frac{1}{(k_3+Q)^2 + m_{\mathsf P}^2} {\cal G}_{\tau\upsilon}^{\mathsf P}(k_3,Q) \nonumber \\
& \quad +
{\cal G}_{\rho\tau}^{\mathsf P}(k_2,k_3) \frac{1}{(k_1+Q)^2 + m_{\mathsf P}^2} {\cal G}_{\nu\upsilon}^{\mathsf P}(k_1,Q) \nonumber \\
& \quad +
{\cal G}_{\tau\nu}^{\mathsf P}(k_3,k_1) \frac{1}{(k_2+Q)^2 + m_{\mathsf P}^2} {\cal G}_{\rho\upsilon}^{\mathsf P}(k_2,Q) , \label{PoleApprox}
\end{align}
where
\begin{equation}
 {\cal G}_{\tau\upsilon}^{\mathsf P}(k,Q) =
 \frac{e^2}{4\pi^2} \varepsilon_{\tau\upsilon\alpha\beta} k_{\alpha} Q_{\beta} G_{\mathsf P}(k^2,Q^2) ,
 \label{TransitionFF}
\end{equation}
etc., are the $\mathsf P \gamma^{(\ast)}\gamma^{(\ast)}$ transition form factors (discussed below),
with the other tensors in Eq.\,\eqref{PoleApprox} being obvious by analogy.
%
%

In writing Eq.\,\eqref{TransitionFF}, we have supposed that the {\sf P}-meson is on-shell, so that
$2 k\cdot Q = - (k^2 + Q^2 + m_P^2)$.
This on-shell approximation delivers an unambiguous pole-contribution to $a_\mu$ that is readily compared across existing calculations.
Following Ref.\,\cite{Colangelo:2015ama}, all off-shell contributions are defined to be expressed elsewhere.
One might amend this scheme by using solutions of inhomogeneous Bethe-Salpeter equations to replace the {\sf P}-meson propagator and associated bound-state amplitudes.
The impact of such an approach has not yet been examined; but, \emph{prima facie}, if both are done correctly, they should be equivalent.

Now using Eqs.\,\eqref{rebuild}, \eqref{PiZero}, one obtains (recall $k_3+k_1+k_2=0$ when $Q\to 0$):
\begin{align}
&(2\pi)^4  \Pi_{(\upsilon)\nu\rho\tau\kappa}(k_1,k_2,k_3,Q)  \nonumber \\
&  \stackrel{Q\to 0}{=}
\frac{e^2G_{\mathsf P}(k_1^2, k_2^2) e^2G_{\mathsf P}(k_3^2,0)}{k_3^2+m_{\mathsf P}^2}
{\cal D}_{\nu\rho\alpha\beta}^{\tau\kappa\alpha^\prime\upsilon}
k_{1\alpha} k_{2\beta} k_{3\alpha^\prime} \nonumber \\
& \quad +
\frac{e^2G_{\mathsf P}(k_2^2, k_3^2) e^2G_{\mathsf P}(k_1^2,0)}{k_1^2+m_{\mathsf P}^2}
{\cal D}_{\rho\tau\alpha\beta}^{\nu\kappa\alpha^\prime\upsilon}
k_{2\alpha} k_{3\beta} k_{1\alpha^\prime} \nonumber \\
& \quad +
\frac{e^2G_{\mathsf P}(k_3^2, k_1^2) e^2G_{\mathsf P}(k_2^2, 0)}{k_2^2+m_{\mathsf P}^2}
{\cal D}_{\tau\nu\alpha\beta}^{\rho\kappa\alpha^\prime\upsilon}
k_{3\alpha} k_{1\beta} k_{2\alpha^\prime} ,
\end{align}
where
we have introduced the double Levi-Civita tensor:
$
{\cal D}_{\nu\rho\alpha\beta}^{\tau\kappa\alpha^\prime\upsilon} =
\varepsilon_{\nu\rho\alpha\beta}
\varepsilon_{\tau\kappa\alpha^\prime\upsilon}$, etc.

Combining the above elements into an expression for $\Gamma_\upsilon^{\gamma\mu}(p,p) $, Eq.\,\eqref{vertex}, and using Eq.\,\eqref{projectors}, one arrives at a result for $a_\mu$ that is expressed as an integral over two spacelike momenta and one angular variable \cite{Jegerlehner:2009ry}; see Eq.\,\eqref{Finalamu}.
A value for the {\sf P}-pole HLbL contributions to the muon anomalous magnetic moment then follows immediately once the $\mathsf P \gamma^{(\ast)}\gamma^{(\ast)}$ transition form factors are known.

\section{$\pi^0$ Transition Form Factors}
\label{SecThree}
%
%
\subsection{Background}
The $ \pi^0(k_1+k_2)\gamma^{(\ast)}(k_1) \gamma^{(\ast)}(k_2)$ transition form factor has long been discussed because it provides the archetypal example of the anomalies that arise in some currents during the process of quantising a field theory \cite{Adler:1969gk, Bell:1969ts, Adler:2004ih}.
Using CSMs, the model-independent character of the $\pi^0\gamma^{(\ast)}\gamma^{(\ast)} $ anomaly constraint and its connection with the dynamics underlying emergent hadron mass (EHM) were first explained in Ref.\,\cite{Roberts:1994hh}.

The $\pi^0\gamma^{(\ast)} \gamma^{(\ast)} $ process is described by a rank-2 tensor:
\begin{subequations}
\label{piTFF}
\begin{align}
{\mathpzc T}_{\nu\rho}^{\pi}(k_1,k_2) & =
T_{\nu\rho}^{\pi}(k_1,k_2) + T_{\rho\nu}^{\pi}(k_2,k_1)  \\
& = 2 T_{\nu\rho}^{\pi}(k_1,k_2) ,
\end{align}
\end{subequations}
where the last equality owes to Bose symmetry.
At leading order (rainbow-ladder, RL, truncation) in the systematic and symmetry-preserving CSM approximation scheme introduced in Refs.\,\cite{Munczek:1994zz, Bender:1996bb}:
\begin{subequations}
\label{TriDiag}
\begin{align}
\nonumber
2 T_{\nu\rho}^\pi& (k_1,k_2) = \tfrac{e^2}{2\pi^2}\, \epsilon_{\nu\rho\alpha\beta}\,k_{1\alpha} k_{2\beta}\, G_\pi(k_1^2, k_2^2)\\
& =: e^2 \epsilon_{\nu\rho\alpha\beta}\,k_{1\alpha} k_{2\beta}\, F_\pi(k_1^2, k_2^2)\\
\nonumber
& =  2 e^2 {\rm tr}_{\rm D} \int_{dq} \,
 S_\ell(q) i\Gamma_\nu(q,q_1)\, S_\ell(q_1) \Gamma_\pi(q_1,q_2) \, \\
& \quad\quad \times S_\ell(q_2) \, i \Gamma_\rho(q_2,q)\,,
\label{anomalytriangle}
\end{align}
\end{subequations}
where $q_{1}=q+k_1$, $q_{2} = q - k_2$.  
Implicit in Eq.\,\eqref{anomalytriangle} is a meson-dependent flavour-structure quark charge factor, \emph{viz}.\ with $N_c$ being the number of colours, 
\begin{equation}
\mbox{\c{o}}_\pi :=N_c (e_u^2 - e_d^2) = 3 (4/9-1/9) = 1 .
\end{equation}

The Abelian anomaly constraint \cite{Adler:1969gk, Bell:1969ts, Roberts:1994hh, Holl:2005vu} entails that, in the chiral limit, \emph{i.e}., when the renormalisation group invariant (RGI) light-quark current mass vanishes,
\begin{equation}
F_\pi(0,0) = \frac{1}{4\pi^2 f_\pi^0} ,
\label{EqABJ}
\end{equation}
where $f_\pi^0 \approx 0.088\,$GeV is the chiral-limit pion leptonic decay constant \cite{Holl:2005vu}.  In any approach with a veracious connection to QCD, Eq.\,\eqref{EqABJ} must emerge in a manner that is independent of any approximation/truncation.  This outcome is a feature of CSMs \cite{Roberts:1994hh, Holl:2005vu}.

At the other kinematic extremes, QCD predicts \cite{Lepage:1980fj}:
\begin{subequations}
\label{QCDlarget}
\begin{align}
\lim_{t\to \infty} t F_\pi(t,0) & = 2 f_\pi \mbox{\c{o}}_\pi = 2 f_\pi =: {\mathpzc n}_\pi^\infty \, \label{UVpit0}\\
\lim_{t\to \infty} t F_\pi(t,t) & = \frac{2}{3} f_\pi \mbox{\c{o}}_\pi  = \frac{2}{3} f_\pi =: {\mathpzc n}_\pi^{\infty_2} . \label{UVpitt}
\end{align}
\end{subequations}
In these limits, too, CSMs deliver parameter-free agree\-ment with the SM predictions \cite{Maris:2002mz, Raya:2015gva, Raya:2016yuj, Ding:2018xwy, Raya:2019dnh}.

\subsection{Calculation}
There are three key elements in Eq.\,\eqref{anomalytriangle}.
The first is the light-quark propagator (we work in the isospin symmetry limit, so $u=\ell=d$):
\begin{equation}
\label{GenSp}
S_\ell(q) = -i \gamma\cdot q \, \sigma_V^\ell(q^2;\zeta^2)+\sigma_S^\ell(q^2;\zeta^2) ,
\end{equation}
which, consistent with Eq.\,\eqref{anomalytriangle}, is obtained from the rainbow-truncation gap equation ($\zeta$ is the renormalisation scale).
For each quark flavour, $f$, the ratio $M^f(q^2)=\sigma_S^f(q^2;\zeta^2)/\sigma_V^f(q^2;\zeta^2)$ is RGI.  Using this, it is common to define an associated RGI Euclidean constituent quark mass, $M_E^f$, as the solution of:
\begin{equation}
M_E^f= M^f([M_E^f]^2) .
\label{ECQM}
\end{equation}

The next is the pion Bethe-Salpeter amplitude ($P=k_1+k_2$) \cite{Maris:1997tm}:
\begin{align}
\nonumber \Gamma_\pi(&\hat k;P)  =  \gamma_5 \left[ i E_\pi(\hat k;P) + \gamma\cdot P \, F_\pi(\hat k;P)  \right.\\
& \left. + \gamma\cdot \hat k \hat k\cdot P \, G_\pi(\hat k;P) + \sigma_{\mu\nu} \hat k_\mu P_\nu\, H_\pi(\hat k;P) \right]\,, 
\label{Gammapi}
\end{align}
calculated in RL truncation.
Here, referring to the meson amplitudes in Eq.\eqref{anomalytriangle}, we have written $q_1=\hat k+\eta P$, $q_2=\hat k-(1-\eta) P$, $\eta\in[0,1]$; and Poincar\'e covariance entails that no observable can depend on $\eta$, i.e., the definition of the relative momentum.
(Since all systems considered herein are charge-conjugation eigenstates, one can, without loss of generality but with a marked gain in simplicity, choose $\eta = 1/2$.  In this case, each function in Eq.\,\eqref{Gammapi} is even under $\hat k\cdot Q \to - \hat k\cdot Q$.)

The third piece is the photon-quark vertex, $\Gamma_\nu(q_i,q_f)$, which should also be calculated in RL truncation.  This Dirac-matrix-valued function has eleven independent tensor structures modulated by nonzero scalar functions of two variables \cite{Ball:1980ay}.

A brief review of the elements required to implement RL truncation is provided in Appendix\,\ref{AppendixRL}.  
Using the information provided therein, one can solve the necessary gap and Bethe-Salpeter equations using what are now standard algorithms \cite{Maris:1997tm, Maris:2005tt, Krassnigg:2009gd}.  

Importantly, when working in Euclidean metric, one has $(k_1+k_2)^2=P^2 < 0$ in Eq.\,\eqref{anomalytriangle}, so each each integrand function is sampled at complex values of its argument.  
This is an issue because RL truncation delivers dressed-quark propagators that possess at least one pair of complex conjugate poles \cite{Maris:1997tm, Windisch:2016iud}.
(The specific analytic structure of QCD dressed propagators can depend (strongly) on the kernel \cite{Burden:1991gd}; nonetheless, confinement is expected to entail a violation of reflection positivity \cite[Sec.\,5]{Ding:2022ows}.)
Consequently, the gap equation can only be solved directly for $S(q)$ within a parabola in the complex $q^2$ plane whose locus is determined by the location of this first pair of poles.  
However, with increasing $k_{1}^2$ and/or $k_{2}^2$, one typically finds that a domain larger than that covered by the parabola is required to complete calculation of the integral.  
One must therefore choose an approach for analytic continuation of either the integrand or the integrated result.

In Ref.\,\cite{Raya:2019dnh}, this was addressed using parametrisations of gap and Bethe-Salpeter equation solutions, built from results obtained within the parabola and expressed via PTIRs.  
These fits possess a user-chosen analytic structure; so, the Eq.\,\eqref{anomalytriangle} integral can be evaluated using the Cauchy integral formula for any selected set of $(k_1^2,k_2^2)$ values.
(To ease computational effort, Ref.\,\cite{Raya:2019dnh} only used this method on the rays $(k^2,0)$ and $(k^2,k^2)$, otherwise developing fits for $F_{\mathsf P}(k_1^2,k_2^2)$, $k_1^2\neq k_2^2 \neq 0$.)
Naturally, the integral's value at each $(k_1^2,k_2^2)$ depends on the developed PTIRs.  
The approach has been validated for elastic pion and kaon form factors. \cite{Yao:2024drm}.

\begin{figure}[t]
\centerline{%
\includegraphics[clip, width=0.48\textwidth]{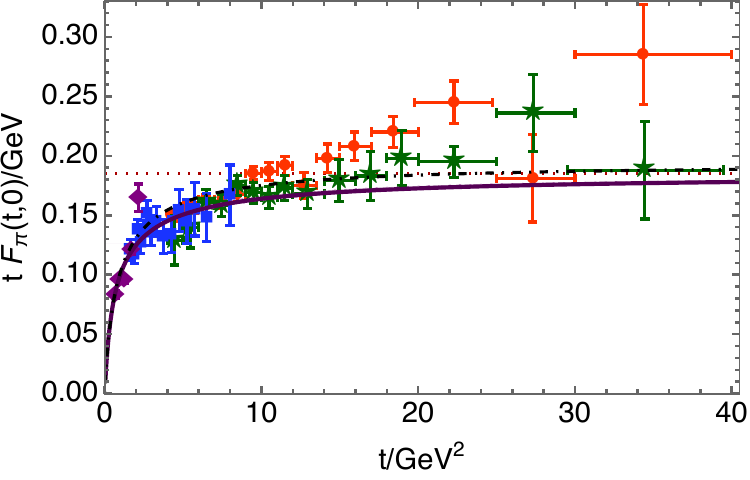}}
\vspace*{-2ex}

\caption{\label{FpiHOS}
Half off-shell $\pi^0$ transition form factor: $t F_\pi(t,0)$.
Legend.
Solid purple curve: result obtained herein, with SPM used to complete the curve on $t>3\,$GeV$^2$.
Dot-dashed black curve: obtained using PTIRs \cite{Raya:2015gva} and used in Ref.\,\cite{Raya:2019dnh}.
Dotted red -- asymptotic limit \cite{Lepage:1980fj}.
Data: BaBar \cite{Aubert:2009mc} -- circles (red); CELLO \cite{Behrend:1990sr} -- diamonds (purple); CLEO \cite{Gronberg:1997fj} -- squares (blue); Belle \cite{Uehara:2012ag} -- stars (green).
}
\end{figure}

To check the Ref.\,\cite{Raya:2019dnh} approach for the $a_\mu$ case, a different scheme is adopted herein.
Namely, for each value of the pairs $(k^2, 0)$ or $(k^2,k^2)$ that ensures the domain of integration remains within the parabola described above, the integrand in Eq.\,\eqref{anomalytriangle} is decomposed into the active Poincar\'e-invariant scalars: 
$q^2$, $q\cdot k_1$, $q\cdot k_2$.  This means the integral to be computed involves one ``radial''  and two angular integrations.
A Gaussian quadrature is then introduced to numerically define the integral.  The quadrature specifies a set of points in the complex plane at which the integrand functions will be evaluated: each point has three coordinates.
At those points, one solves for the integrand functions and thereby arrives at the following tables of numbers:
two three-dimensional (3D) arrays for the light-quark propagator (two functions of complex-valued arguments);
four 3D arrays for the four functions in the pion Bethe-Salpeter amplitude;
and eleven 3D arrays for the photon--light-quark vertex -- seventeen arrays altogether.
Following this procedure, interpolations are not needed and one directly obtains a result for the integral by calculating the weighted sum associated with the Gaussian quadrature.  
The procedure is repeated until the answer is independent of the discretisation grid: 
typically, 40 radial points and 24 points for each angle. 
Importantly, the integral in Eq.\,\eqref{anomalytriangle} is convergent, so one quickly loses sensitivity to the numerical ultraviolet regularisation parameter. 

\begin{table}[t]
\caption{ \label{FPvaluesCAparams}
Predictions for pseudoscalar meson transition form factors at the kinematic limits: $(t_1,t_2)=(0,0), (\infty,0), (\infty,\infty)$ -- columns 1\,-\,3;
and associated Canterbury approximant parameters -- see Eqs.\,\eqref{Canterbury1}, \eqref{Canterbury2} -- needed to complete $F_{\mathsf P}(t_1,t_2)$ -- columns 4\,-\,7.
(Units.  
First column $=$ $1/{\rm GeV}$; 
columns 2, 3 $=$ ${\rm GeV}$; 
column 4 $=$ $1/{\rm GeV}^2$; 
columns 5, 7 $=$ $1/{\rm GeV}^4$;
column 6 $=$ $1/{\rm GeV}^6$.)
}
\begin{tabular*}
{\hsize}
{
l@{\extracolsep{0ptplus1fil}}
|c@{\extracolsep{0ptplus1fil}}
c@{\extracolsep{0ptplus1fil}}
c@{\extracolsep{0ptplus1fil}}
|c@{\extracolsep{0ptplus1fil}}
c@{\extracolsep{0ptplus1fil}}
c@{\extracolsep{0ptplus1fil}}
c@{\extracolsep{0ptplus1fil}}}\hline\hline
\centering
    & $F_{\mathsf P}\ $ & ${\mathpzc n}_{\mathsf P}^\infty\ $ & ${\mathpzc n}_{\mathsf P}^{\infty_2} \ $ & $d_{01}\ $ & $d_{10}\ $ & $d_{11}\ $ & $d_{20}\ $ \\\hline
 %
        $\pi^0\ $ & 0.274$\phantom{00}\ $ & 0.185$\ $ & 0.0615$\ $ & 6.440$\ $ & 2.444$\ $ & 1.645\phantom{0}$\ $& 0.742\phantom{000}$\ $ \\ 
$\eta\ $ & 0.260$\phantom{00}\ $ & 0.144$\ $ & 0.0478$\ $  & $5.497\ $  & $2.577\ $ & $0.804\phantom{0}\ $ & $0.586\phantom{000}\ $ \\
%
$\eta^\prime\ $ & $0.321\phantom{00}\ $ & $0.298\ $ & $0.0994\ $ & $3.071\ $ & $2.019\ $ & $0.508\phantom{0}\ $ & $0.322\phantom{000}\ $ \\
%
$\eta_c\ $ & $0.0540\phantom{0}\ $ & $0.747\ $ & $0.249\phantom{0}\ $ & $3.219\ $ & $1.559\ $ & $0.327\phantom{0}\ $ & $0.127\phantom{000}\ $ \\
%
$\eta_b\ $ & $0.00286\ $ & $0.373\ $ & $0.124\phantom{0}\ $ & $0.084\ $ & $0.014\ $ & $0.0009\ $  & $0.000014\ $\\
\hline
\end{tabular*}
\end{table}

\subsection{Numerical results: $\pi^0$}
\subsubsection{Half off shell -- $\pi^0$}
\label{piHOS}
Our prediction for the half off-shell pion transition form factor, $F_\pi(t,0)$ is drawn in Fig.\,\ref{FpiHOS}, with the result for $F_\pi(0,0)$ given in Table~\ref{FPvaluesCAparams}: we predict a roughly 5\% decrease from the chiral limit value in Eq.\,\eqref{EqABJ}.  
Our value yields the following $\pi^0 \to \gamma\gamma$ decay width ($\alpha$ is the fine structure constant):
\begin{equation}
\Gamma_{\pi^0\gamma\gamma} = \tfrac{\pi}{4}\alpha^2 m_{\pi^0}^3 F_\pi(0,0)^2 = 7.75\,{\rm eV} ,
\label{widthform}
\end{equation}
which may be compared with the experimental value  \cite{Gasparian:2016oyl}: $7.74(13)\,$eV.
This calculation used the moving frame Bethe-Salpeter amplitude; see Appendix~\ref{AppendixMoving}.

On $t\in [0,3]\,$GeV$^2$, $F_\pi(t,0)$ is obtained by direct evaluation of the integral in Eq.\,\eqref{anomalytriangle}.  
However, on $t> 3\,$GeV$^2$, the integrand functions are required on a domain that extends beyond the parabola of analyticity.  In this case, we employ the Schlessinger point method (SPM) \cite{Schlessinger:1966zz, PhysRev.167.1411, Tripolt:2016cya} to obtain the result by extrapolation.  
The SPM is now commonly used in such cases, to good effect; for recent examples, see, \textit{e.g}., Refs.\,\cite{Cui:2022dcm, Yao:2024drm, Yao:2024ixu, Cheng:2024cxk, Xu:2025ups}.

The SPM is a tool from applied mathematics that enables the reliable use of multipoint Pad\'e approximants in the interpolation and extrapolation of analytic functions.
It is able to reliably reconstruct a function in the complex plane within a radius of convergence determined by that one of the function's branch points located nearest to the real domain from which the sample points are drawn.  
Furthermore, modern implementations introduce a statistical element; consequently, all extrapolations come with an objective and reliable estimate of uncertainty.  Hence, we use the SPM to obtain predictions for $F_\pi(t,0)$ on the domain $t > t_\pi^{\rm S}:= 3\,$GeV$^2$.

We implement the SPM as follows.
\begin{description}
\item[Step 1] For $F_\pi(t,0)$, we produce $N=40$ directly calculated values of $t F_\pi(t,0)$, spaced evenly on $Q^2 \leq t_\pi^{\rm S}$.

\item [Step 2]  From that set, $M_0=8$ points are chosen at random.  
A $9$-parameter continued-fraction interpolation of those points is then constructed, with the ninth (deepest embedded) coefficient fixed by requiring that the $t \to \infty$ extrapolation of the interpolation function reproduces Eq.\,\eqref{UVpit0}.
(With an odd number of continued-fraction coefficients, one obtains an approximant that approaches a constant as $t\to\infty$.)
%
%
The interpolant is accepted if it is a function that is both singularity-free and monotonically increasing on $t\leq 1\,000\,$GeV$^2$.

\item[Step 3]
Step 2 is repeated with another set of $M_0$ randomly chosen points.
(More than 75-million independent sets of points can be chosen in this way.)

\item[Step 4]
One continues with 2 and 3 until $n_{M_0}=1000\,$ acceptable interpolations/extrapolations are obtained, at which point one has $1\,000$ statistically independent representations of $t F_\pi(t,0)$.

\end{description}
Working with this set of interpolations/extrapolations, one obtains the prediction for $tF_\pi(t,0)$ drawn in Fig.\,\ref{FpiHOS} (solid purple curve).  The bracketing uncertainty band indicates the domain covered by 68\% of the extrapolations, i.e., it is the $1\sigma$ uncertainty band. 
Having implemented Eq.\,\eqref{UVpit0} as a constraint, the band is practically invisible. 
Evidently, insofar as comparisons with data are concerned, our \textit{ab initio} RL prediction and the result obtained in Ref.\,\cite{Raya:2015gva}, using PTIRs and an \textit{Ansatz} for the photon-quark vertex, are not meaningfully different.

\subsubsection{Equally off shell -- $\pi^0$}
\begin{figure}[t]
\centerline{%
\includegraphics[clip, width=0.48\textwidth]{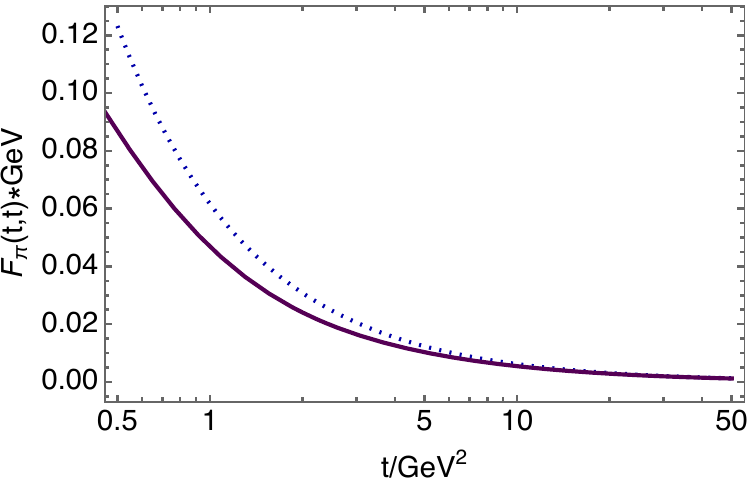}}
\vspace*{-2ex}

\caption{\label{EoffShellpi}
Equally off-shell $\pi^0$ transition form factor: $F_\pi(t,t)$.
Legend.
Solid purple curve: result obtained herein by direct numerical integration. 
Dotted blue curve -- asymptotic behaviour, $(2/3) f_\pi/t$ \cite{Lepage:1980fj}.
}
\end{figure}

Owing to the balanced kinematics involved when calculating $F_\pi(t,t)$, the integrand functions in Eq.\,\eqref{anomalytriangle} are never sampled outside the parabolic domain of analyticity; hence, this transition form factor can be calculated directly to (arbitrarily) large values of $t$ \cite{Maris:2002mz}.
Our prediction is drawn in Fig.\,\ref{EoffShellpi}.
Evidently, alike with Ref.\cite{Maris:2002mz}, on $t \gtrsim 10\,$GeV$^2$, our prediction is indistinguishable from the QCD asymptotic result, Eq.\,\eqref{UVpitt}.

\subsubsection{Off shell, unbalanced momenta -- $\pi^0$}
Consider $F_\pi(t_1,t_2)$.  
When increasing $|t_1 -t_2|$ away from zero, one quickly finds that direct numerical integration forces sampling of the Eq.\,\eqref{anomalytriangle} integrand functions outside the parabolic domain of analyticity.
To overcome the problem, one might imagine using the SPM to construct a representation of the function on its two-dimensional domain of support.  However, that scheme is too cumbersome for a function of two variables.  
Instead, following Ref.\,\cite{Raya:2019dnh}, we choose to employ a Canterbury approximant \cite{canterbury}.

{\allowdisplaybreaks
Within that scheme, we develop the following approximation to $F_\pi(t_1,t_2)$:
\begin{subequations}
\label{Canterbury1}
\begin{align}
F_\pi^{\rm C}(t_1,t_2) & = N(t_1,t_2)/D(t_1,t_2) , \\
N(t_1,t_2) & = n_{00} + n_{10} (t_1+t_2) + n_{01} t_1 t_2  , \\
D(t_1,t_2) & = 1+ d_{10}(t_1+t_2) + d_{01} t_1 t_2 \nonumber \\
& \quad + d_{11} (t_1+t_2) t_1 t_2 + d_{20} (t_1^2 + t_2^2) . 
\end{align}
\end{subequations}
Three of the coefficients are fixed by the known function value at $(t_1,t_2)=(0,0)$ and the asymptotic limits, Eq.\,\eqref{QCDlarget}:
\begin{subequations}
\label{Canterbury2}
\begin{align}
n_{00} & = F_\pi(0,0) , \\
n_{01} & = (2/3) {\mathpzc n}_\pi^\infty d_{11}, \\
n_{10} & = d_{20}  {\mathpzc n}_\pi^\infty . \label{n10d20}
\end{align}
\end{subequations}
The remaining four are determined by requiring that they minimise a least-squares difference between $F_\pi^{\rm C}(t_1,t_2) $ and a dense set of $F_\pi(t,0)$, $F_\pi(t,t)$ values calculated on the domains that bridge the asymptotic limits.  The resulting parameters are listed in Table~\ref{FPvaluesCAparams}.  
Where comparison is possible, they deliver a result for $F_\pi^{\rm C}(t_1,t_2)$ that is practically indistinguishable from the target functions.  Namely, the curves overlap or are indistinguishable within plotting linewidth; stated differently, in all cases, the relative $\mathsf L_1$-difference between the Canterbury approximant and the target function is less than $0.5$\%. 
} 


\begin{table}[t]
\caption{ \label{anommm}
Pseudoscalar meson pole HLbL contributions to the muon anomalous magnetic moment. 
Unweighted means of results reported in Ref.\,\cite{Estrada:2024cfy} are (units $10^{-11}$):
$\pi^0$, $62.0(2.4)$;
$\eta$, $14.8(2.6)$;
$\eta^\prime$, $14.3(2.0)$.}
\begin{tabular*}
{\hsize}
{
l@{\extracolsep{0ptplus1fil}}
|c@{\extracolsep{0ptplus1fil}}
c@{\extracolsep{0ptplus1fil}}
c@{\extracolsep{0ptplus1fil}}}\hline\hline
\centering
${\mathsf P}\times 10^{11}\ $   & herein$\phantom{000}\ $ & Ref.\,\cite{Raya:2019dnh}$\phantom{00000}\ $ & Ref.\,\cite{Gerardin:2023naa}$ \ $  \\\hline
$\pi^0\  $ & $58.2(0.7)\phantom{00000}\ $ & $61.4(2.1) \phantom{00000}\ $  & $57.8(2.0)\ $ \\
$\eta\  $ & $13.1(0.4)\phantom{00000}\ $  & $14.7(1.9) \phantom{00000}\ $  & $11.6(2.0)\ $ \\
$\eta^\prime\  $ & $13.0(0.1)\phantom{00000}\ $ & $13.6(0.8)\phantom{00000}\ $   & $15.7(4.3)\ $ 
\\
$\eta_c\  $ & $\phantom{1}0.824(036)\phantom{000}\ $ & $\phantom{0}0.9(0.1)\phantom{00000} \ $   & $\ $ \\
$\eta_b \ $ & $0.0023(0001) \ $ & $0.0026(0001)\ $ & 
\\
\hline
total$\ $ & $85.1(0.8)\phantom{00000}\ $ & $90.6(2.9)\phantom{00000}\ $ & \\
\hline\hline
\end{tabular*}
\end{table}

\subsection{$\pi^0$ contribution to $a_\mu$}
\label{pi0amu}
Referring to Eq.\,\eqref{Finalamu}, it is plain that we now have all inputs necessary to calculate the $\pi^0$ pole HLbL contribution to the muon anomalous magnetic moment.  Our prediction is listed in Table~\ref{anommm}.

Regarding Eq.\,\eqref{Finalamu}, uncertainties in our prediction stem from the results we have for $F_\pi(t,0)$, $F_\pi(t_1,t_2)$.
As noted above, for all intents and purposes, the Canterbury approximant reproduces the target functions precisely.  
Hence, the only true underlying source of uncertainty is that in the SPM result for $F_\pi(t,0)$.  
A basic SPM uncertainty is obtained by using the Canterbury approximant, $F_\pi^C(t_1,t_2)$, determined by the SPM mean value for $F_\pi(t,0)$, calculating $a_\mu^{HLbL_{\pi^0}}$ via Eq.\,\eqref{Finalamu} using each one of the $1\,000$ approximants for $F_\pi(t,0)$, 
then combining the results.
Naturally, one could also have developed Canterbury approximants for each one of the $1\,000$ SPM results for $F_\pi(t,0)$.  
This would deliver an additional uncertainty correlated with that arising from $F_\pi(t,0)$, but it would not lead to magnification by two because there is no SPM uncertainty in $F_\pi(t,t)$.
To be conservative, therefore, we list results with the basic SPM uncertainty multiplied by two.  This is an upper bound.

\section{$\eta$, $\eta^\prime$ Transition Form Factors}
\label{etaprimeeta}
\subsection{Non-Abelian anomaly}
\begin{figure}[t]
\centerline{%
\includegraphics[clip, width=0.46\textwidth]{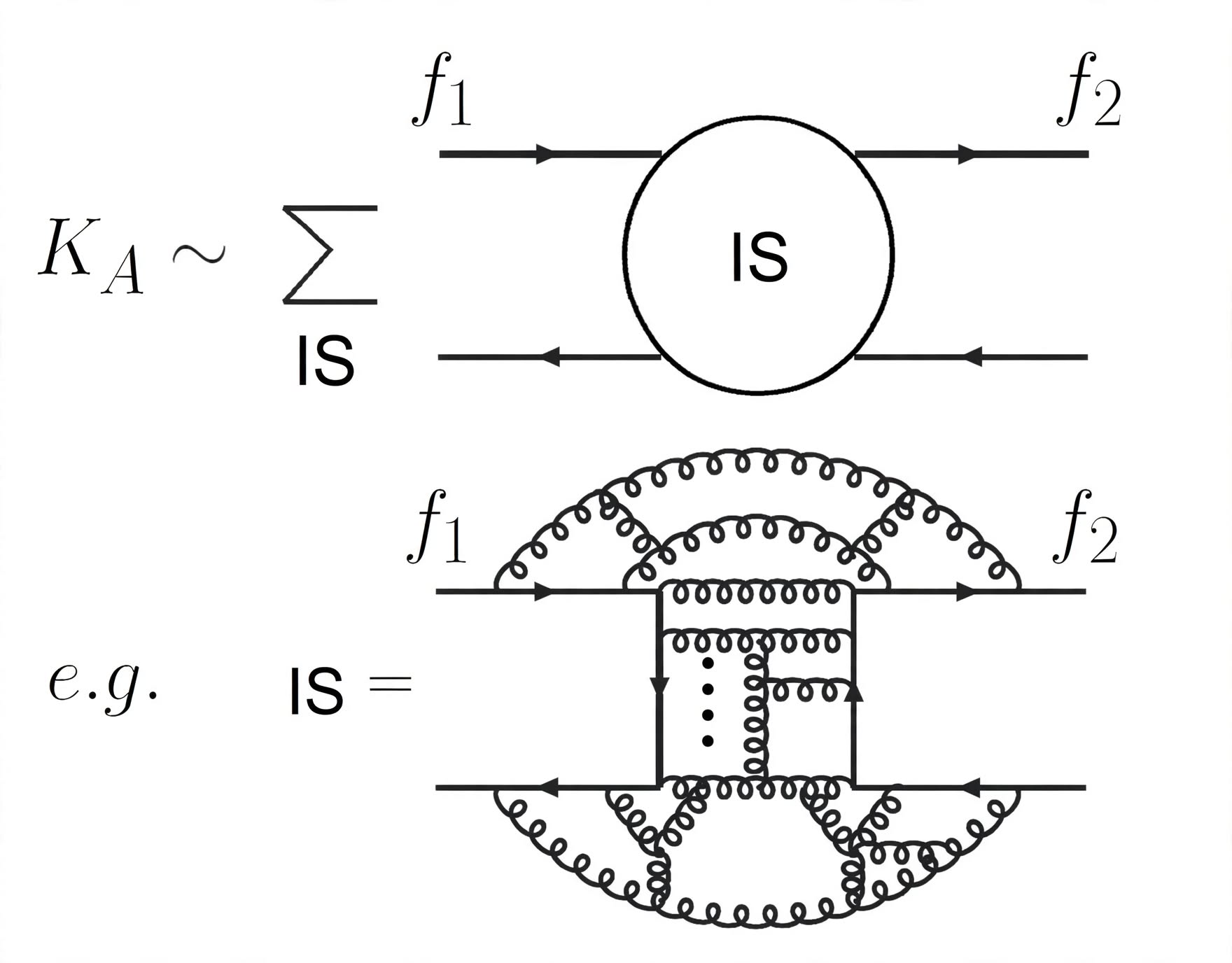}}
\vspace*{-2ex}

\caption{\label{FigAnomaly}
Contributions to the Bethe-Salpeter kernel deriving from the non-Abelian anomaly must have the sort of ``hairpin'' structure depicted here; and any intermediate state (\textsf{IS}) must involve infinitely many lines.  (Straight lines denote quarks, with $f_1$ and $f_2$ independent, and springs denote gluons.)}
\end{figure}

These two mesons are special because the properties of each are significantly influenced by the non-Abelian anomaly \cite{Christos:1984tu}; hence, RL truncation cannot work.  A CSM approach to this problem is presented in Ref.\,\cite{Ding:2018xwy}.  It includes the additional ``anomaly'' contribution to the Bethe-Salpeter kernel sketched in Fig.\,\ref{FigAnomaly}.  We follow and update the Ref.\,\cite{Ding:2018xwy} treatment herein.  (In the limit of perfect isospin symmetry, which we employ, the $\pi^0$ does not mix with $\eta$, $\eta^\prime$.)

Considering the character of the non-Abelian anomaly, it becomes clear that its contribution to the Bethe-Salpeter kernel cannot contain external quark or antiquark lines which are simply connected to the internal lines.
Instead, purely gluonic configurations (potentially not representable diagrammatically) must mediate, as sket\-ched in Fig.\,\ref{FigAnomaly}.  

Consequently, in the $\eta, \eta^\prime$ channels, the Bethe-Salpeter kernel can be decomposed into a sum:
\begin{equation}
\label{EqGenK}
{\mathpzc K} = {\mathpzc K}_{N} + {\mathpzc K}_A\,.
\end{equation}
Here, ${\mathpzc K}_N$ is that part which can be constructed diagrammatically and involves all those contributions that are accessible in perturbation theory and therefore contribute at ultraviolet momenta.  This is where the RL kernel sits.
The other term, ${\mathpzc K}_A$, is the non-Abelian anomaly contribution, indicated in Fig.\,\ref{FigAnomaly}: it is essentially nonperturbative and hence possesses material support only at infrared momenta.

The RL part of the Eq.\,\eqref{EqGenK} kernel is detailed in Appendix~\ref{AppendixRL}.
For ${\mathpzc K}_A$, we adopt the form discussed in Ref.\,\cite[Sec.\,II.D.3]{Ding:2018xwy}.
This \textit{Ansatz} is expressed in terms of three parameters.
(1) $D_\xi$: the overall strength of the Gaussian whose momentum dependence defines the profile of the anomaly interaction.  The Gaussian range is the same as that of the infrared part of the RL kernel, \textit{viz}.\ $\omega$ in Eq.\,\eqref{defcalG}.
(2) $\theta_\xi$: the relative strength of pseudoscalar and axialvector components of ${\mathpzc K}_A$. 
(3) $r_A$: the strength of U$(3)$-flavour symmetry breaking in the hairpin structure of Fig.\,\ref{FigAnomaly}.

In solving the Bethe-Salpeter equations, it is common to adopt a $U(N_f=3)$ quark flavour basis \cite{Feldmann:1998vh, Feldmann:1998sh}, wherewith the associated Bethe-Salpeter amplitudes can be written $(\ell=u=d)$:
\begin{subequations}
\begin{align}
\Gamma_{\eta,\eta^\prime}&(q;P)  = \mathbb{F}^\ell \Gamma^\ell_{\eta,\eta^\prime}(q;P) + \mathbb{F}^s \Gamma_{\eta,\eta^\prime}^s(q;P)\,,
\label{chiFB}\\
\mathbb{F}^\ell & =
\left(\begin{array}{ccc}
1 & 0 & 0 \\
0 & 1 & 0 \\
0 & 0 & 0 \\
\end{array}\right)\,,\quad
\mathbb{F}^s  =
\left(\begin{array}{ccc}
0 & 0 & 0 \\
0 & 0 & 0 \\
0 & 0 & \surd 2 \\
\end{array}\right)\,.
\end{align}
\end{subequations}
The Dirac-matrix-valued coefficients $\Gamma^{\ell,s}_{\eta,\eta^\prime}(q;P)$ in Eq.\,\eqref{chiFB} are Bethe-Salpeter amplitudes which, respectively, describe the momentum-space $\ell\bar \ell$ or $s \bar s$ correlations in the $\eta$, $\eta^\prime$: $q$ is the relative momentum between the valence-quarks and $P$ is the bound-state's total momentum.
Each such amplitude has the standard four-component form -- see Eq.\,\eqref{Gammapi}, wherein, hereafter, without loss of generality, we choose $\eta=1/2$. 

Using canonically normalised solutions of the Bethe-Salpeter equations, one can calculate the flavour-projected $\eta$, $\eta^\prime$ leptonic decay constants ($q_\pm =q \pm P/2$):
\begin{align}
f_{\eta,\eta^\prime}^{\ell,s}P_\mu & = Z_2\,N_c  {\rm tr}_{\rm D}\int_{dq}^\Lambda \gamma_5\gamma_\mu \nonumber \\
& \quad \times 
S_{\ell,s}(q_+) \Gamma_{\eta,\eta^\prime}^{\ell,s}(q;P)S_{\ell,s}(q_-)\,,
\label{eqfM}
\end{align}
where $\int_{dq}^\Lambda$ indicates a symmetry-preserving regularisation of the four-dimensional integral.
Owing to the multiplicative factor, $Z_2$, which is the quark wave function renormalisation constant, the right-hand side is independent of both the regularisation scale and renormalisation point \cite{Maris:1997tm}.

At this point, we solve the Bethe-Salpeter equations to obtain $\eta$, $\eta^\prime$ masses and amplitudes, use those results to calculate the leptonic decay constants in Eq.\,\eqref{eqfM}, and vary $D_\xi$, $\theta_\xi$, $r_A$ so as to obtain a least-squares best fit to empirical $\eta$, $\eta^\prime$ masses \cite[PDG]{ParticleDataGroup:2024cfk} and phenomenologically inferred leptonic decay constants  \cite{Feldmann:1998sh, Benayoun:1999au, DeFazio:2000my}.
The best fit parameters and results are listed in Table~\ref{currentmass}.

In discussing the analogues of Eq.\,\eqref{QCDlarget} for $\eta$, $\eta^\prime$, it is useful to introduce U$(3)$ flavour-octet and -singlet axial-vector currents \cite{Leutwyler:1997yr}:
\begin{equation}
\left(
\begin{array}{cc}
f_\eta^8 & f_\eta^0\\
f_{\eta^\prime}^8 & f_{\eta^\prime}^0\\
\end{array}\right)
=
\left(\begin{array}{cc}
f_8 \cos\theta_8 & -f_0\sin\theta_0\\
f_8 \sin\theta_8 & \phantom{-}f_0\cos\theta_0\\
\end{array}\right).
\label{OSmixing}
\end{equation}
In this case, flavour projection is implemented using the usual U$(3)$ generators in the fundamental representation.   

The merit of the octet/singlet basis is that the octet axial-vector current satisfies a standard Ward-Green-Takahashi identity; so, $f_8$ in Eq.\,\eqref{OSmixing} is independent of the renormalisation scale, $\zeta$.  
In contrast, the flavour-singlet axialvector current is anomalous in QCD.  
Consequently, $f_0$ and, thus, $f^{l,s}_{\eta,\eta^\prime}$ in Eq.\,\eqref{eqfM}, depend on $\zeta$.  
Their decrease with $\zeta$ is not rapid, however, because the leading contribution to the anomalous dimension is O$(\alpha^2)$, where $\alpha$ is the QCD running coupling \cite{Leutwyler:1997yr, Feldmann:1997vc, Feldmann:1998sh, Agaev:2014wna}.  

In these terms, one has $\mathsf P= \eta, \eta^\prime$ ($\tau = \Lambda_{\rm QCD}^2/t$):
\begin{subequations}
\label{UVzeta}
\begin{align}
t F_{\mathsf P}(t,0) & \stackrel{\tau \simeq 0}{\approx}
\left. 6 \left[ c_\ell f_{\mathsf P}^\ell(t) + c_s f_{\mathsf P}^s(t)\right] \right|_{\tau=0}\\
& = \left. 2 \left[ c_8 f_{\mathsf P}^8 + 2 c_0 f_{\mathsf P}^0(t) \right]\right|_{\tau=0}
=: {\mathpzc n}_{\mathsf P}^\infty ,
\end{align}
\end{subequations}
$c_\ell = 5/9$, $c_s = \surd 2/9$, 
$c_8 = 1/\surd 3$, $c_0=\sqrt{2/3}$, where \cite{Agaev:2014wna}
\begin{equation}
\label{f0zeta}
f_M^0(\zeta^2) = f_M^0(\zeta_2^2) \left( 1 + \frac{2 n_f}{\pi\beta_0} \left[\alpha_S(\zeta^2) - \alpha_S(\zeta_2^2)\right]\right)\,,
\end{equation}
with $\alpha_S$ being the strong running coupling.
Recall, we renormalise at $\zeta_2$.
In analogy with Eq.\,\eqref{UVpitt}:
\begin{equation}
{\mathpzc n}_{\eta,\eta^\prime}^{\infty_2} := \left. [t F_{\eta,\eta^\prime}(t,t)]\right|_{\tau \simeq 0}
= \frac{1}{3} {\mathpzc n}_{\eta,\eta^\prime}^\infty .
\label{UVpitteta}
\end{equation}

Equation~\eqref{f0zeta} presents the formula for leading-order evolution of the flavour-singlet leptonic decay constant.  In connection with evolution of $\eta$, $\eta^\prime$ transition form factors, a complete procedure is described elsewhere \cite[Sec.\,IV.B]{Ding:2018xwy}.  We employ that scheme herein.  For practical purposes, it implements the following mappings ($t_{12}=t_1+t_2$):
\begin{subequations}
\label{evolutionNAA}
\begin{align}
F_{\mathsf P}(t_1,& t_2) 
\to \left\{ 
\begin{array}{ll}
F_{\mathsf P}(t_1,t_2) & t_{12} \leq 2{\rm GeV}^2 \\ 
F_{\mathsf P}(t_1,t_2) {\mathpzc E}_{\mathsf P}(t_{12}) & t_{12}> 2{\rm GeV}^2 \\ 
\end{array}
\right. \\
{\mathpzc E}_{\eta} & = 
\frac{0.24 +1.042 t_{12} + 0.614 t_{12}^2}{0.24 + 0.255 t_{12} + 1.242 t_{12}^2} \nonumber \\
& \quad \times 
\frac{1 + 8.355 t_{12} + 7.136 t_{12}^2 + 7.330 t_{12}^3}
{1+ 14.91 t_{12} + 6.179 t_{12}^2 + 4.003 t_{12}^3} , 
\\
%
%
{\mathpzc E}_{\eta^\prime} &  =
\frac{1 + 0.677 t_{12} + 0.01012 t_{12}^2}{0.189+ 0.00316 t_{12}} \nonumber \\
& \quad \times 
\frac{0.355 + 15.41 t_{12} + 12.85 t_{12}^2}{1 + 85.86 t_{12} + 119.9 t_{12}^2 + 45.18 t_{12}^3} .
\end{align}
\end{subequations}

\subsection{Triangle diagram}
As with the neutral pion, $\gamma^{(\ast)}\gamma^{(\ast)} \to \eta,\eta^\prime$ transitions are each described by a single form factor, which, in RL truncation, is obtained from the following expression:
\begin{subequations}
\label{TransitionFFs}
\begin{align}
\nonumber
& \epsilon_{\nu\rho\alpha\beta}\,k_{1\alpha} k_{2\beta}\, F_{\mathsf P}(k_1^2, k_2^2)\\
\nonumber
& = \tfrac{1}{2\pi^2}\,  \epsilon_{\nu\rho\alpha\beta}\,k_{1\alpha} k_{2\beta}\, G_{\mathsf P}(k_1^2, k_2^2)\\
& = \tfrac{1}{2\pi^2}\,  \epsilon_{\nu\rho\alpha\beta}\,k_{1\alpha} k_{2\beta}\, 
[G_{\mathsf P}^\ell(k_1^2, k_2^2) + G_{\mathsf P}^s(k_1^2, k_2^2)] \\
& = 2 N_c {\rm tr}_{\rm D}\int_{dq}\bigg[
c_\ell i \chi_\nu^\ell(q,q_1) \Gamma_{\mathsf P}^\ell(q_1,q_2) S_\ell(q_2) i \Gamma_\rho^\ell (q_2,q) 
\nonumber \\
& \qquad+ c_s i\chi_\nu^s(q,q_1) \Gamma_{\mathsf P}^s(q_1,q_2) S_s(q_2) i\Gamma_\rho^s(q_2,q)\bigg]\,,
\label{etaIntegral}
\end{align}
\end{subequations}
where
$\chi_\nu^{\ell,s}(q,q_1) = S_{\ell,s}(q) \Gamma_\nu^{\ell,s}(q,q_1) S_{\ell,s}(q_1)$ 
and 
$(k_1+k_2)^2 = -m_{\mathsf P}^2$.

\subsection{Numerical results: $\eta$}
\subsubsection{Half off shell}
\begin{figure}[t]
\centerline{%
\includegraphics[clip, width=0.48\textwidth]{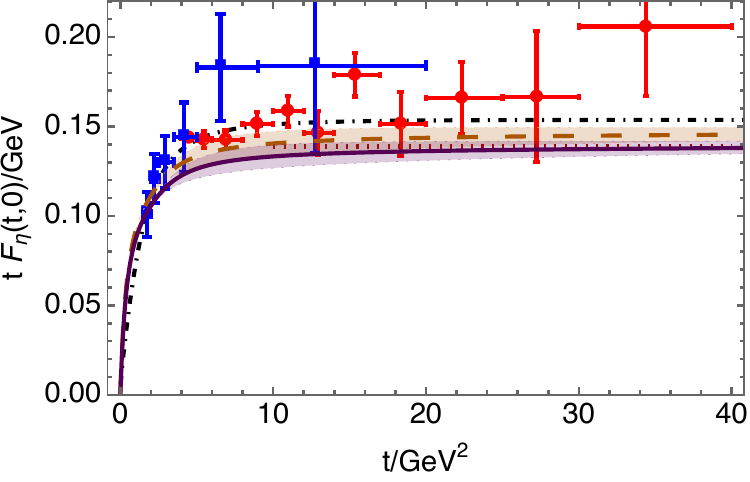}}
\vspace*{-2ex}

\caption{\label{FigetaHOS}
Half off-shell $\eta$ transition form factor: $t F_\eta(t,0)$.
Legend.
Solid purple curve: result obtained herein, with SPM used to complete the curve on $t>2.5\,$GeV$^2$.
Long-dashed orange curve: obtained from purple curve and band rescaled ($+5$\%) to recover the empirical value of $\Gamma_{\eta \gamma \gamma} $.
Dot-dashed black curve: obtained using PTIRs \cite{Raya:2015gva} and used in Ref.\,\cite{Raya:2019dnh}.
Dotted red -- asymptotic limit, obtained using Eq.\,\eqref{UVzeta} with evolution as described in Ref.\,\cite{Ding:2018xwy}.
Data.
CLEO \cite{Gronberg:1997fj} -- squares (blue); 
BaBar \cite{BABAR:2011ad} -- circles (red). 
}
\end{figure}

Our prediction for the half off-shell $\eta$ transition form factor, $F_\eta(t,0)$, is drawn in Fig.\,\ref{FigetaHOS}, with the result for $F_\eta(0,0)$ given in Table~\ref{FPvaluesCAparams}.
Adapting Eq.\,\eqref{widthform}, one finds
\begin{equation}
\Gamma_{\eta \gamma \gamma} = 0.464\,{\rm keV} , 
\label{etawidth}
\end{equation}
which may be compared with the experimental value \cite{ParticleDataGroup:2024cfk}: $0.516(22)\,$keV.
Agreement with this value would be achieved if $F_\eta(0,0)$ were increased by 5\%.  
(It is worth noting that in working with transition form factor results from Ref.\,\cite{Ding:2018xwy}, Ref.\,\cite{Raya:2019dnh} was using a parameter to fit the value of  $F_\eta(0,0)$ and hence $\Gamma_{\eta \gamma \gamma}$.)
Our result for $F_\eta(t,0)$ was obtained using the moving frame Bethe-Salpeter amplitude; see Appendix~\ref{AppendixMoving}.

As with the pion case, the integral in Eq.\,\eqref{etaIntegral} can only be calculated directly on a limited domain, \textit{viz}.\  for the $\eta$, $t<2.5 $GeV$^2$.  We complete the $F_\eta(t,0)$ curve by employing precisely the same SPM procedure described in connection with $F_\pi(t,0)$.  

Regarding Fig.\,\ref{FigetaHOS}, it is apparent that our prediction for $F_\eta(t,0)$ differs somewhat from the PTIR result used in Ref.\,\cite{Raya:2019dnh} and lies below the data.  A somewhat different picture is obtained if one rescales our prediction by the factor necessary to recover the empirical value of  $\Gamma_{\eta \gamma \gamma} $.  We will bear this in mind when reporting a value for the $\eta$-meson contribution to $a_\mu$.

\subsubsection{Equally off shell -- $\eta$}
\begin{figure}[t]
\centerline{%
\includegraphics[clip, width=0.48\textwidth]{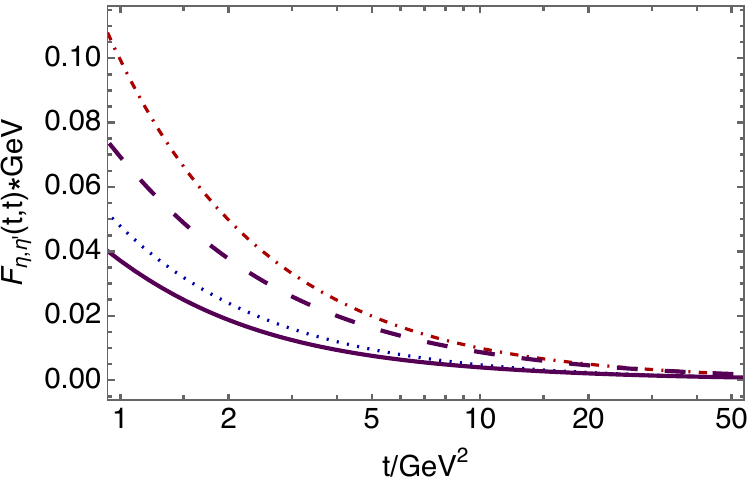}}
\vspace*{-2ex}

\caption{\label{EoffShelleta}
Equally off-shell $\eta$, $\eta^\prime$ transition form factors: $F_{\eta,\eta^\prime}(t,t)$.
Legend.
Long-dashed purple curve: $\eta^\prime$ result obtained herein by direct numerical integration. 
Dot-dashed red curve -- asymptotic behaviour, ${\mathpzc n}_{\eta^\prime}^{\infty_2}/t$ with ${\mathpzc n}_{\eta^\prime}^{\infty_2}$ from Table~\ref{FPvaluesCAparams}.
Solid purple curve: $\eta$ result obtained herein by direct numerical integration. 
Dotted blue curve -- asymptotic behaviour, ${\mathpzc n}_\eta^{\infty_2}/t$ with ${\mathpzc n}_\eta^{\infty_2}$ from Table~\ref{FPvaluesCAparams}.
}
\end{figure}

Again, owing to the balanced kinematics, the integrand functions in Eq.\,\eqref{anomalytriangle} for $F_\eta(t,t)$ are not sampled outside the parabolic domain of analyticity; hence, this transition form factor can be calculated directly for all values of $t$.
Our prediction is drawn in Fig.\,\ref{EoffShelleta}.
Evidently, alike with the $\pi^0$ case, on $t \gtrsim 10\,$GeV$^2$, our prediction is indistinguishable from the QCD asymptotic result, Eq.\,\eqref{UVpitteta}.

\subsubsection{Off shell, unbalanced momenta -- $\eta$}
Alike with $F_\pi(t_1,t_2)$, one finds that $F_\eta(t_1,t_2)$ can only be directly calculated on narrow strips around $|t_1 -t_2|=0$.  
In the $\eta$ case, 
we populate a measurable domain in the neighbourhood of the ray $\{(t_1,t_2) \, | \, 0\lesssim t_1/{\rm GeV}^2 < 2.5, t_2= 0\}$; 
use the calculated result for $F_\eta(t,t)$; 
capitalise on the infrared and ultraviolet asymptotic limits of the transition form factor;
and with this information, build a Canterbury approximant as described in connection with Eqs.\,\eqref{Canterbury1}, \eqref{Canterbury2}.  The resulting parameters are listed in Table~\ref{FPvaluesCAparams}.  
Where comparison is possible, they deliver a result for $F_{\eta}^{\rm C}(t_1,t_2)$ that is practically indistinguishable from the target functions.  Namely, the curves overlap or are indistinguishable within plotting linewidth; stated differently, in all cases, the relative $\mathsf L_1$-difference between the Canterbury approximant and the target function is $\lesssim 1.0$\%.  


\subsubsection{$\eta$ contribution to $a_\mu$}
With the inputs now available, we can complete the calculation of the $\eta$ pole HLbL contribution to the muon anomalous magnetic moment.  Our prediction is listed in Table~\ref{anommm}.  
The recorded value includes the effects of evolution -- Eq.\,\eqref{evolutionNAA}, 
magnification --  discussed in connection with Eq.\,\eqref{etawidth} and Fig.\,\ref{FigetaHOS};
and the uncertainty is multiplied by a factor of two, as was done for the $\pi^0$ -- Sec.\,\ref{pi0amu}.
The unamended result is $12.5(2)$.

\begin{figure}[t]
\centerline{%
\includegraphics[clip, width=0.48\textwidth]{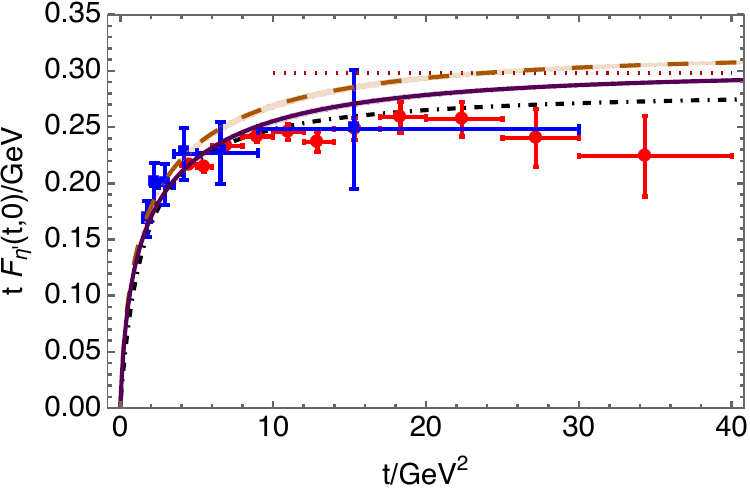}}
\vspace*{-2ex}

\caption{\label{FigetaprimeHOS}
Half off-shell $\eta'$ transition form factor: $t F_{\eta^\prime}(t,0)$.
Legend.
Solid purple curve: result obtained herein, with SPM used to complete the curve on $t>1.0\,$GeV$^2$.
Long-dashed orange curve: obtained from purple curve and band rescaled ($+5$\%) to recover a  value of $\Gamma_{\eta^\prime \gamma \gamma} $ that is consistent with experiment.
Dot-dashed black curve: obtained using PTIRs \cite{Raya:2015gva} and used in Ref.\,\cite{Raya:2019dnh}.
Dotted red -- asymptotic limit, obtained using Eq.\,\eqref{UVzeta} with evolution as described in Ref.\,\cite{Ding:2018xwy}.
Data.
CLEO \cite{Gronberg:1997fj} -- squares (blue); 
BaBar \cite{BABAR:2011ad} -- circles (red). 
}
\end{figure}

\subsection{Numerical results: $\eta^\prime$}
\subsubsection{Half off shell}
Our prediction for the half off-shell $\eta^\prime$ transition form factor, $F_{\eta^\prime}(t,0)$, is drawn in Fig.\,\ref{FigetaprimeHOS}, with the result for $F_{\eta^\prime}(0,0)$ given in Table~\ref{FPvaluesCAparams}.
Adapting Eq.\,\eqref{widthform}, one finds
\begin{equation}
\Gamma_{\eta^\prime \gamma \gamma} = 3.79\,{\rm keV} , 
\label{etaPwidth}
\end{equation}
which may be compared with the experimental value \cite{ParticleDataGroup:2024cfk}: $4.34(20)\,$keV.
Using the $5$\% rescaling factor determined from the $\eta$ analysis, one obtains 
$\Gamma_{\eta^\prime \gamma \gamma} = 4.21$\,{\rm keV}, in line with experiment.  
(Again, recall that this matches the approach followed in Ref\,\cite{Raya:2019dnh}, which, practically, used a single parameter to fit the $\eta$, $\eta^\prime$ two-photon decay widths.)
Our result for $F_{\eta^\prime}(t,0)$ was obtained using the rest frame Bethe-Salpeter amplitude; see Appendix~\ref{AppendixRest}.  

Alike with the $\pi^0$, $\eta$ cases, the integral in Eq.\,\eqref{etaIntegral} can only be calculated directly on a limited domain, \textit{viz}.\  for the $\eta^\prime$, $t<1.0 $\,GeV$^2$.  We complete the $F_{\eta^\prime}(t,0)$ curve by employing the SPM procedure described in connection with $F_\pi(t,0)$.  

Comparing Figs.\,\ref{FigetaHOS} and \ref{FigetaprimeHOS}, it is apparent that the SPM uncertainty on the $\eta$ result is somewhat larger than that associated with the $\eta^\prime$ curve.  
This is because one must employ the moving frame for the $\eta$ calculation, in which case solving the Bethe-Salpeter equation is a more computationally intensive exercise than when using the rest frame.  
Consequently, with a given integration quadrature, one obtains a less precise Bethe-Salpeter amplitude.   
Notwithstanding, since the uncertainty on $a_\mu^\eta$ is $\lesssim 3$\%, we choose to accept the balance of computational time and precision  that we have achieved. 

Looking at Fig.\,\ref{FigetaprimeHOS}, one sees that our prediction for $F_{\eta^\prime}(t,0)$ lies above the PTIR result used in Ref.\,\cite{Raya:2019dnh}.
It also traces a trajectory above the extant largest-$t$ data, but those data express an unexpected (unphysical) trend (falling in magnitude) and sit not much more than $1\sigma$ away from our prediction. 
The $5$\% rescaled curve sits somewhat higher still.  We will consider these things when delivering a value for the $\eta^\prime$-meson contribution to $a_\mu$.

\subsubsection{Equally off shell -- $\eta^\prime$}
As will now be familiar, in cases with balanced kinematics, the integrand functions in Eq.\,\eqref{anomalytriangle} are not sampled outside the parabolic domain of analyticity.
Hence, $F_{\eta^\prime}(t,t)$ can be calculated directly for all values of $t$.
Our prediction is drawn in Fig.\,\ref{EoffShelleta}: on $t \gtrsim 10\,$GeV$^2$, it is indistinguishable from the QCD asymptotic result, Eq.\,\eqref{UVpitteta}.

\subsubsection{Off shell, unbalanced momenta -- $\eta^\prime$}
Similar to the cases discussed above, $F_{\eta^\prime}(t_1,t_2)$ can only be directly calculated on narrow strips around $|t_1 -t_2|=0$.  
So, for the $\eta^\prime$, 
we populate a measurable domain in the neighbourhood of the ray $\{(t_1,t_2) \, | \, 0\lesssim t_1/{\rm GeV}^2 < 1, t_2= 0\}$; 
use the calculated result for $F_{\eta^\prime}(t,t)$; 
capitalise on the infrared and ultraviolet asymptotic limits of the transition form factor;
and with this information, build a Canterbury approximant as described in connection with Eqs.\,\eqref{Canterbury1}, \eqref{Canterbury2}.  The resulting parameters are listed in Table~\ref{FPvaluesCAparams}.  
Alike with the above cases, where comparison is possible, they deliver a result for $F_{\eta^\prime}^{\rm C}(t_1,t_2)$ that is practically indistinguishable from the target functions: the curves overlap or are indistinguishable within plotting linewidth.
Stated differently, in all cases, the relative $\mathsf L_1$-difference between the Canterbury approximant and the $\eta^\prime$ target function is $\lesssim 0.3$\%.  


\subsubsection{$\eta^\prime$ contribution to $a_\mu$}
%
Our prediction for the $\eta^\prime$ HLbL contribution to the muon anomalous magnetic moment is listed in Table~\ref{anommm}.  
The reported value includes the effects of evolution -- Eq.\,\eqref{evolutionNAA}, 
magnification --  discussed in connection with Eq.\,\eqref{etaPwidth} and Fig.\,\ref{FigetaprimeHOS};
and the uncertainty is multiplied by a factor of two, as was done for the $\pi^0$ -- Sec.\,\ref{pi0amu}.
The unamended result is $12.33(03)$.

\section{$\eta_c$ Transition Form Factors}
\label{Secetac}
In principle, the $\eta_c$ transition form factor is uncomplicated.  One need simply evaluate the integral in Eq.\,\eqref{anomalytriangle} using propagators and vertices appropriate to this meson.  In this case, the flavour-structure quark charge factor is 
\begin{equation}
\mbox{\c{o}}_{\eta_c} :=  
N_c e_c^2  = \frac{4}{3} . 
\end{equation}
Consequently, 
\begin{subequations}
\label{QCDlargetetac}
\begin{align}
\lim_{t\to \infty} t F_{\eta_c}(t,0) & = 2 f_{\eta_c} \mbox{\c{o}}_{\eta_c} = \tfrac{8}{3} f_{\eta_c} =: {\mathpzc n}_{\eta_c}^\infty , \label{UVetact0}\\
\lim_{t\to \infty} t F_{\eta_c}(t,t) & = \tfrac{2}{3} f_{\eta_c} \mbox{\c{o}}_{\eta_c}  = \tfrac{8}{9} f_{\eta_c} =: {\mathpzc n}_{\eta_c}^{\infty_2} , \label{UVetactt}
\end{align}
\end{subequations}
where $f_{\eta_c}$ is the leptonic decay constant of the $\eta_c$ meson; see Table~\ref{currentmass}.  
The rates at which these ultraviolet limits are approached are discussed below. 

\subsection{Numerical results: $\eta_c$}
\begin{figure}[t]
\centerline{%
\includegraphics[clip, width=0.48\textwidth]{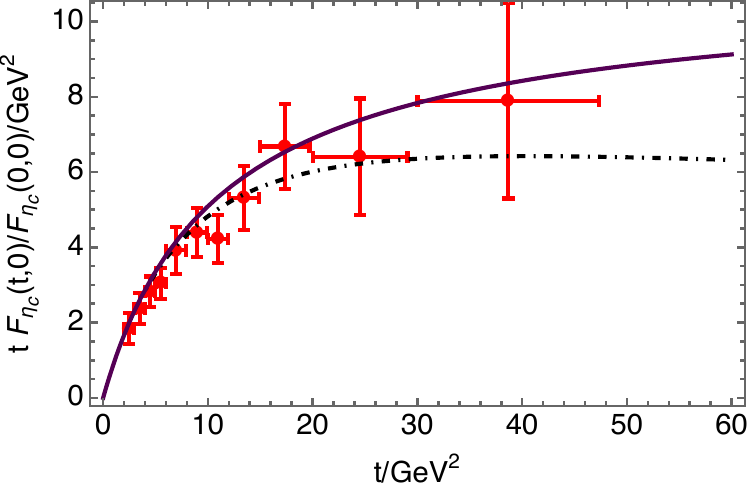}}

\centerline{%
\includegraphics[clip, width=0.48\textwidth]{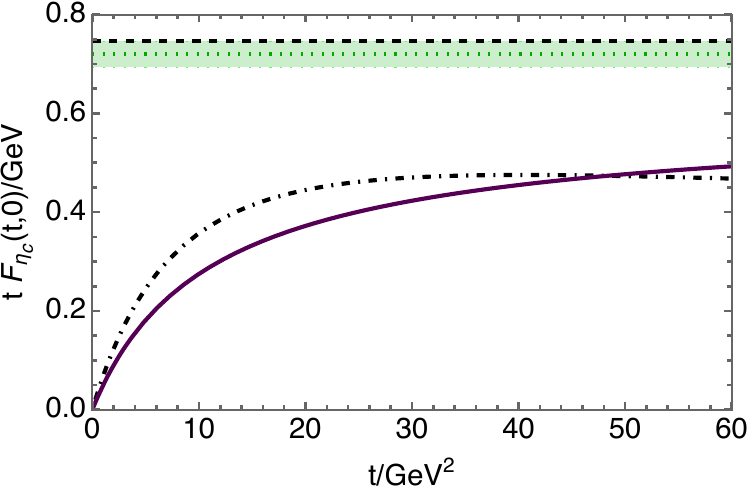}}

\vspace*{-2ex}

\caption{\label{FigetacHOS}
{\sf Panel A}.
Normalised half off-shell $\eta_c$ transition form factor: $t F_{\eta_c}(t,0)/F_{\eta_c}(0,0)$.
Legend.
Solid purple curve: result obtained herein.  (The SPM is used to complete the curve on $t>9\,$GeV$^2$.  The associated uncertainty is indiscernible.)
Dot-dashed black curve: obtained using PTIRs \cite{Raya:2016yuj} and used in Ref.\,\cite{Raya:2019dnh}.
%
Data.
BaBar \cite{Lees:2010deBaBar} -- circles (red). 
{\sf Panel B}.
Approach of $t F_{\eta_c}(t,0)$ toward the ultraviolet limit in Eq.\,\eqref{UVetact0}.
Legend.
Solid purple curve: result obtained herein. 
Dot-dashed black curve: obtained using PTIRs \cite{Raya:2016yuj} and used in Ref.\,\cite{Raya:2019dnh}.
Dashed blue curve -- Eq.\,\eqref{UVetact0} evaluated using $f_{\eta_c}$ from Table~\ref{currentmass}; and 
dotted green curve and associated band -- same but using the empirical result for $f_{\eta_c}$ in Table~\ref{currentmass}.
}
\end{figure}

\subsubsection{Half off shell}
Our prediction for the half off-shell $\eta_c$ transition form factor, $F_{\eta_c}(t,0)$, is drawn in Fig.\,\ref{FigetacHOS}A, with the result for $F_{\eta_c}(0,0)$ given in Table~\ref{FPvaluesCAparams}.  
The curve was obtained using the rest frame Bethe-Salpeter amplitude; see Appendix~\ref{AppendixRest}.  
Adapting Eq.\,\eqref{widthform}, one has
\begin{equation}
\Gamma_{\eta_c \gamma \gamma}  = 
\tfrac{\pi}{4}\alpha^2 m_{\eta_c}^3 F_{\eta_c}(0,0)^2 = 3.24 \,{\rm keV} ,
\label{etacwidth}
\end{equation}
which may be compared with the experimental value \cite{ParticleDataGroup:2024cfk}: $6.4(5)\,$keV.
In order to reproduce the empirical value, one would need to increase $F_{\eta_c}(0,0)$ by $41$\%.  Following Ref.\,\cite{Raya:2016yuj}, the Ref.\,\cite{Raya:2019dnh} study achieved this match by tuning a parameter in the photon--$c$-quark vertex \emph{Ansatz}.
We will bear these things in mind when reporting a result for the $\eta_c$ contribution to $a_\mu$.

\begin{figure}[t]
\centerline{%
\includegraphics[clip, width=0.48\textwidth]{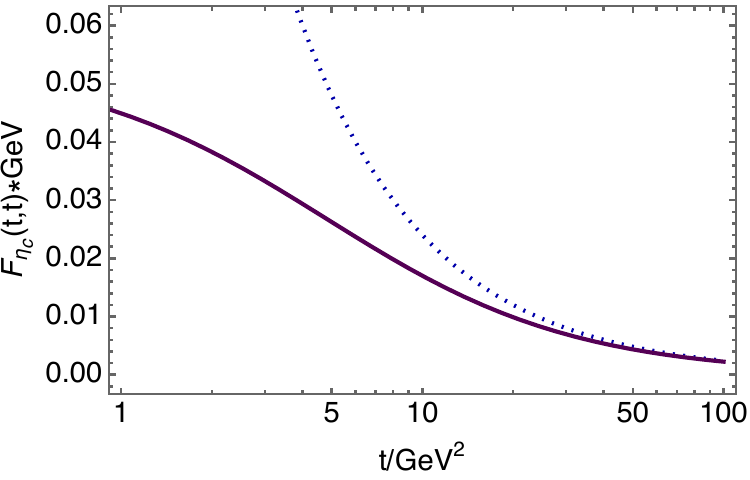}}
\vspace*{-2ex}

\caption{\label{EoffShelletac}
Equally off-shell $\eta_c$ transition form factor: $F_{\eta_c}(t,t)$.
Legend.
Solid purple curve: result obtained herein by direct numerical integration. 
Dotted red curve -- asymptotic behaviour, $(8/9) f_{\eta_c}/t$, where $f_{\eta_c}$ is our predicted value for this leptonic decay constant; see Table~\ref{currentmass}.
}
\end{figure}

For the $\eta_c$, the integral in Eq.\,\eqref{anomalytriangle} can be calculated directly on a (large but) limited domain, \textit{viz}.\ $t \lesssim 9.0 $GeV$^2$.  
The upper bound is much bigger than is possible for lighter mesons.
We complete the $F_{\eta_c}(t,0)$ curve by employing the SPM procedure, much as described in connection with $F_\pi(t,0)$; see Sec.\,\ref{piHOS}.

There is a difference, however, which arises because the $c$-quark is neither light nor truly heavy and is made necessary by the outcome highlighted in Fig.\,\ref{FigetacHOS}B.
Namely, as observed in Ref.\,\cite{Raya:2016yuj}, 
$t F_{\eta_c}(t,0)$ approaches its ultraviolet limit, Eq.\,\eqref{UVetact0}, very slowly 
because $m_{\eta_c}$ is large and, consequently, the associated distribution amplitude long remains measurably far from the QCD asymptotic form.
Moreover, unlike in the lighter meson cases discussed above, the mass-retarded, logarithmic character of the slow approach cannot be mimicked adequately by the rational polynomials used in SPM analyses.  
So, it is not reasonable to fix the ninth SPM coefficient via Eq.\,\eqref{UVetact0}; see {\bf Step 2} in Sec.\,\ref{piHOS}.
Indeed, if one attempts to do so, then one finds practically zero SPM interpolators that satisfy the constraints imposed in {\bf Step 2}.

Therefore, we steadily decreased the ultraviolet limit value until the same number of randomly selected SPM interpolators were obtained as in the pion case ($40$\%): this required
\begin{equation}
{\mathpzc n}_{\eta_c}^\infty \to 0.8 \, {\mathpzc n}_{\eta_c}^\infty .
\label{UVetact0New}
\end{equation} 
It was with this value that we completed the SPM analysis of $F_{\eta_c}(t,0)$, under the assumption that, on some domain well beyond that actually sampled by the integral in Eq.\,\eqref{anomalytriangle}, this function logarithmically achieves its true asymptotic limit.

\subsubsection{Equally off shell -- $\eta_c$}
Once again, in this case of balanced kinematics, the integrand functions in Eq.\,\eqref{anomalytriangle} are not sampled outside the parabolic domain of analyticity; hence, one can directly calculate $F_{\eta_c}(t,t)$ for all values of $t$.
Our prediction is drawn in Fig.\,\ref{EoffShelletac}: on $t \gtrsim 50\,$GeV$^2$, it is indistinguishable from the QCD asymptotic result, Eq.\,\eqref{UVetactt}.
Comparing with Fig.\,\ref{EoffShelleta}, one sees that, for the $\eta_c$, the asymptotic trajectory is recovered at a larger value of $t$ than it is for the $\eta^\prime$; nevertheless, the onset delay is better measured by $m_{\eta_c}/m_{\eta^\prime}$ than the square of this ratio; so, it is still reached fairly quickly. 

\subsubsection{Off shell, unbalanced momenta -- $\eta_c$}
As usual, it is necessary to develop a Canterbury approximant for $F_{\eta_c}(t_1,t_2)$. 
Owing to the issue discussed in connection with Fig.\,\ref{FigetacHOS}, \emph{viz}.\  slow approach to the ultraviolet limit by $F_{\eta_c}(t,0)$, Ref.\,\cite{Raya:2019dnh} used a simpler approach in this case.  
We find, instead, that a Canterbury approximant is readily developed, using the now standard method, so long as one modifies Eq.\,\eqref{n10d20} to account for Eq.\,\eqref{UVetact0New}.
The resulting parameters are listed in Table~\ref{FPvaluesCAparams}.  
Alike with the above cases, where comparison is possible, they deliver a result for $F_{\eta_c}^{\rm C}(t_1,t_2)$ that is practically indistinguishable from the target functions: the relative $\mathsf L_1$-difference between the Canterbury approximant and the $\eta_c$ target function is $< 0.5$\%.  

\subsection{$\eta_c$ contribution to $a_\mu$}
%
Our prediction for the $\eta_c$ pole HLbL contribution to the muon anomalous magnetic moment is listed in Table~\ref{anommm}.  
In obtaining the reported value, we considered the impact of our initial underestimate of $\Gamma_{\eta_c \gamma \gamma} $.  
To correct for that, we introduced the following enhancement factor:
\begin{equation}
{\rm mag}^{\eta_c}(t_{12}) = 1+ 0.41/[1+t_{12}/(M_E^c)^2]\,,
\end{equation}
with the range of this magnification fixed by the $c$-quark Euclidean constituent mass, and multiplied ${\rm mag}^{\eta_c}(t_{12}) $ into both $F_{\eta_c}$ functions that appear in Eq.\,\eqref{Finalamu}.  We estimated the uncertainty by changing $M_E^c \to M_E^c (1 \pm 0.1)$ and then multiplied the result by two, as was done for the $\pi^0$ -- Sec.\,\ref{pi0amu}.  
The unamended result is $0.62 \times 10^{-11}$.
In correcting for the $\eta$, $\eta^\prime$ width underestimates, we simply used a constant enhancement factor.  This is not reliable for the $\eta_c$ because the $\eta_c$ transition form factor is significantly stiffer, i.e., damps much more slowly: see, e.g., Fig.\,\ref{FigetacHOS} cf.\  Figs.\,\ref{FigetaHOS}, \ref{FigetaprimeHOS}.

It is worth highlighting that $a_\mu^{\eta_c}$ is larger than the uncertainty on all other HLbL contributions to $a_\mu$ considered hitherto; so, it should not be neglected.

\section{$\eta_b$ Transition Form Factors}
\label{Secetab}
Since the $b$-quark is truly heavy, calculation of the $\eta_b$ transition form factor is uncomplicated.  One need only evaluate the integral in Eq.\,\eqref{anomalytriangle} using propagators and vertices appropriate to this meson.  In this case, the flavour-structure quark charge factor is 
\begin{equation}
\mbox{\c{o}}_{\eta_b} :=  
N_c e_b^2  = \frac{1}{3} . 
\end{equation}
Consequently, 
\begin{subequations}
\label{QCDlargetetab}
\begin{align}
\lim_{t\to \infty} t F_{\eta_b}(t,0) & = 2 f_{\eta_b} \mbox{\c{o}}_{\eta_b} = \tfrac{2}{3} f_{\eta_b} =: {\mathpzc n}_{\eta_b}^\infty , \label{UVetabt0}\\
\lim_{t\to \infty} t F_{\eta_b}(t,t) & = \tfrac{2}{3} f_{\eta_b} \mbox{\c{o}}_{\eta_c}  = \tfrac{2}{9} f_{\eta_b} =: {\mathpzc n}_{\eta_b}^{\infty_2} , \label{UVetabtt}
\end{align}
\end{subequations}
where $f_{\eta_b}$ is the leptonic decay constant of the $\eta_b$ meson; see Table~\ref{currentmass}.  

\subsection{Numerical results: $\eta_b$}
\subsubsection{Half off shell}
\begin{figure}[t]
\centerline{%
\includegraphics[clip, width=0.48\textwidth]{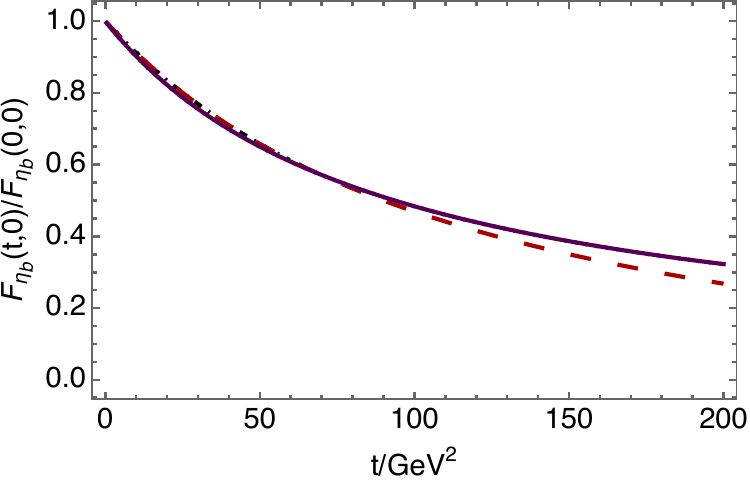}}
\vspace*{-2ex}

\caption{\label{FigetabHOS}
Normalised half off-shell $\eta_b$ transition form factor: $F_{\eta_b}(t,0)/F_{\eta_b}(0,0)$.
Legend.
Solid purple curve: result obtained herein.  (The SPM is used to complete the curve on $t>81\,$GeV$^2$.  Since the direct calculation is available on a large domain, the associated uncertainty is indiscernible.)
Dot-dashed black curve on $0<t/{\rm GeV}^2< 60$: obtained using PTIRs \cite{Raya:2016yuj} and used in Ref.\,\cite{Raya:2019dnh}.
Long-dashed red curve: interpolation and extrapolation of the PTIR result. 
}
\end{figure}

Our prediction for the half off-shell $\eta_b$ transition form factor, $F_{\eta_b}(t,0)$, is drawn in Fig.\,\ref{FigetabHOS}, with the result for $F_{\eta_b}(0,0)$ given in Table~\ref{FPvaluesCAparams}.  
The curve was obtained using the rest frame Bethe-Salpeter amplitude; see Appendix~\ref{AppendixRest}.  

The decay $\eta_b \to \gamma \gamma$ has not yet been observed \cite{ParticleDataGroup:2024cfk}.  
However, adapting the standard quantum mechanics expression for $^1\!S_0 \to \gamma\gamma$ decays, one has
\begin{equation}
\Gamma_{\eta_b \gamma\gamma} = 96 \pi \alpha^2 e_b^4  \frac{f_{\eta_b}^2}{12 m_{\eta_b}} 
= 0.55\,{\rm keV} ,
\end{equation}
using $f_{\eta_b}$ from Table~\ref{currentmass}.  Alternatively, adapting Eq.\,\eqref{widthform}, one has
\begin{equation}
\Gamma_{\eta_b \gamma \gamma}  = 
\tfrac{\pi}{4}\alpha^2 m_{\eta_b}^3 F_{\eta_b}(0,0)^2 = 0.28 \,{\rm keV} .
\label{etabwidth}
\end{equation}
In order to align these two values, one would need to increase $F_{\eta_b}(0,0)$ by $40$\%, i.e., practically the same magnification needed for $F_{\eta_c}(0,0)$.   
In this case, too, following Ref.\,\cite{Raya:2016yuj}, the Ref.\,\cite{Raya:2019dnh} study matched the decay width by tuning a parameter in the photon--$b$-quark vertex \emph{Ansatz}.
We will keep these things in mind when reporting a result for the $\eta_b$ contribution to $a_\mu$.

For the $\eta_b$, the integral in Eq.\,\eqref{anomalytriangle} can be calculated directly on a large domain, \textit{viz}.\ $t \lesssim 81 $GeV$^2$.  
So, in this case, it is straightforward to complete the $F_{\eta_b}(t,0)$ curve by employing the SPM procedure, precisely as described in connection with $F_\pi(t,0)$; see Sec.\,\ref{piHOS}.  In the $9$-parameter continued-fraction interpolations, one can fix the ninth (deepest embedded) coefficient by requiring that Eq.\,\eqref{UVetabt0} is satisfied and still find that 25\% of the randomly generated interpolant curves are both singularity-free and monotonically increasing on $t\leq 1\,000\,$GeV$^2$.

\begin{figure}[t]
\centerline{%
\includegraphics[clip, width=0.48\textwidth]{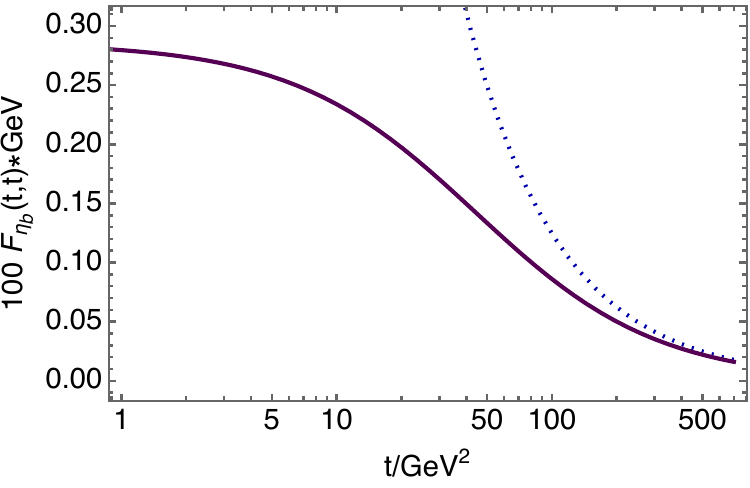}}
\vspace*{-2ex}

\caption{\label{EoffShelletab}
Equally off-shell $\eta_b$ transition form factor: $F_{\eta_b}(t,t)$.
Legend.
Solid purple curve: result obtained herein by direct numerical integration. 
Dotted blue curve -- asymptotic behaviour, $(2/9) f_{\eta_b}/t$, where $f_{\eta_b}$ is our predicted value for this leptonic decay constant; see Table~\ref{currentmass}.
}
\end{figure}

\subsubsection{Equally off shell -- $\eta_b$}
Our prediction for $F_{\eta_b}(t,t)$ is drawn in Fig.\,\ref{EoffShelletab}: on $t \gtrsim 500\,$GeV$^2$, our prediction is indistinguishable from the QCD asymptotic result, Eq.\,\eqref{UVetabtt}.
Comparing with Fig.\,\ref{EoffShelleta}, one sees that, for the $\eta_b$, the asymptotic trajectory is recovered at a much larger value of $t$ than it is for the $\eta^\prime$.
For this truly heavy meson, the onset delay is well measured by $(m_{\eta_b}/m_{\eta^\prime})^2$.

\subsubsection{Off shell, unbalanced momenta -- $\eta_b$}
As in the preceding cases, it is necessary to develop a Canterbury approximant for $F_{\eta_b}(t_1,t_2)$. 
For the $\eta_b$, that is straightforward.  One can follow the same procedure as employed for the pion, implementing the Eq.\,\eqref{QCDlargetetab}-based analogues of Eq.\,\eqref{Canterbury2}.
The resulting parameters are listed in Table~\ref{FPvaluesCAparams}.  
Alike with the above cases, where comparisons are possible, they deliver a result for\linebreak $F_{\eta_b}^{\rm C}(t_1,t_2)$ that is practically indistinguishable from the target functions (linewidth overlap).  In this case, the relative $\mathsf L_1$-difference between the Canterbury approximant and the $\eta_b$ target function is $\lesssim 3.5$\%, measured on the (extensive) domain $0 \leq  t \leq 10\,000\,$GeV$^2$.

\subsection{$\eta_b$ contribution to $a_\mu$}
Our prediction for the $\eta_b$ HLbL contribution to the muon anomalous magnetic moment, $a_\mu$, is listed in Table~\ref{anommm}.  
In obtaining the reported value, we considered the impact of our initial underestimate of $\Gamma_{\eta_b \gamma \gamma} $.  
To correct for that, we introduced the following enhancement factor:
\begin{equation}
{\rm mag}^{\eta_b}(t_{12}) = 1+ 0.40/[1+t_{12}/(M_E^b)^2]\,,
\end{equation}
with the range of this magnification fixed by the $b$-quark Euclidean constituent mass, and multiplied ${\rm mag}^{\eta_b}(t_{12}) $ into both $F_{\eta_b}$ functions that appear in Eq.\,\eqref{Finalamu}.  We estimated the uncertainty by changing $M_E^b \to M_E^b (1 \pm 0.1)$ and then multiplied the result by two, as was done for the $\pi^0$ -- Sec.\,\ref{pi0amu}.  
The unamended result is $0.00172 \times 10^{-11}$.

Evidently, $a_\mu^{\eta_b}$ is too small to have a measurable impact on the sum of pseudoscalar meson pole HLbL contributions to $a_\mu$, i.e., it can be neglected. 

\section{Summary and Perspective}
\label{epilogue}
This is not the first estimate of the pseudoscalar meson pole hadron light-by-light (HLbL) contribution to the muon anomalous magnetic moment obtained using continuum Schwinger function methods (CSMs); but it is the first to entirely employ an internally consistent rainbow-ladder truncation (plus non-Abelian anomaly) treatment of the full five meson problem.  
Moreover, where extrapolations are necessary to complete the calculation of a required pseudoscalar meson transition form factor, it is novel in exploiting the Schlessinger point method, therewith delivering an objective (function form/practitioner independent) result on the required momentum transfer domain with a quantitatively reliable uncertainty.
With these novel features, our study provides both a valuable check on extant analyses and an improvement thereof.

Our final estimate of the pseudoscalar meson pole HLbL contribution to the muon anomalous magnetic moment is obtained by combining the results in Table~\ref{anommm}:
\begin{equation}
\sum_{\mathsf P=\pi^0, \eta, \eta^\prime, \eta_c, \eta_b}  a_\mu^{{\rm HLbL}_{\mathsf P}} = 
85.1(0.8) \times 10^{-11} .
\end{equation}
It is listed in the final row of Table~\ref{anommm}.  The uncertainty is $\lesssim 1$\%.  

In comparison with the only other study to have hitherto considered the full array of known pseudoscalar mesons \cite{Raya:2019dnh}, our estimate for the HLbL contribution is lower, but the difference is less than $2\sigma$.  Notably, the previous study also employed continuum Schwinger function methods.  However, it used physics-constrained \emph{Ans\"atze} for all elements in the calculation; hence, its approach to the problem is very different from ours.  Since the two studies are independent and fairly consistent, it is reasonable to combine them.  The uncertainty weighted average of the results from these two analyses is:
\begin{equation}
\sum_{\mathsf P=\pi^0, \eta, \eta^\prime, \eta_c, \eta_b}^{\rm average}  a_\mu^{{\rm HLbL}_{\mathsf P}} = 
85.5(1.4) \times 10^{-11} .
\end{equation}
(Here and hereafter uncertainties are rescaled following Ref.\,\cite{Birge:1932hlj}.)

Focusing on the light meson only contribution, we find:
\begin{equation}
\sum_{\mathsf P=\pi^0, \eta, \eta^\prime}  a_\mu^{{\rm HLbL}_{\mathsf P}} = 84.3(0.8) \times 10^{-11} .
\label{CCLM}
\end{equation}
Evidently, the $\eta_c$ contribution is a noticeable correction.

In connection with Eq.\,\eqref{CCLM}, we note that an estimate of the light meson contribution is also available from lattice-regularised QCD \cite{Gerardin:2023naa}: $85.1(5.1) \times 10^{-11}$.

\begin{figure}[t]
\centerline{%
\includegraphics[clip, width=0.48\textwidth]{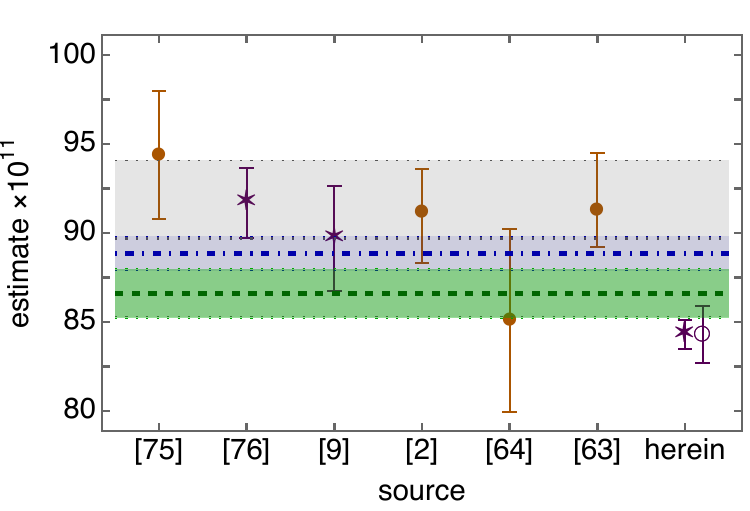}}

%
\caption{\label{AllEstimates}
Contemporary estimates of $\sum_{\mathsf P=\pi^0, \eta, \eta^\prime}  a_\mu^{{\rm HLbL}_{\mathsf P}} $.
Legend.
Entries from left-to-right: Refs.\,\cite{Masjuan:2017tvw}, \cite{Eichmann:2019tjk},  \cite{Raya:2019dnh}, \cite{Aliberti:2025beg}, \cite{Gerardin:2023naa}, \cite{Estrada:2024cfy}, 
with Eq.\,\eqref{CCLM} in the rightmost position.  Left-to-right is also the chronological ordering of estimates.
CSM results, \emph{viz}.\ \cite{Eichmann:2019tjk},  \cite{Raya:2019dnh}, herein: purple crosses.
Dashed green curve within like-coloured band: uncertainty weighted average of the plotted estimates, Eq.\,\eqref{UWA}.
Dot-dashed blue curve and associated band: uncertainty weighted average of the estimates after doubling the uncertainty on our result, Eq.\,\eqref{IUWA}, indicated by the open circle at ``herein''.
Dotted grey line and like-coloured band: unweighted mean of the scattered plotted points.
}
\end{figure}

Furthermore, Ref.\,\cite{Estrada:2024cfy} reports five other independent estimates of the light meson pole component of $a_\mu$.  Together, they are plotted in Fig.\,\ref{AllEstimates} along with our result, Eq.\,\eqref{CCLM}.
The uncertainty weighted average of these estimates is:
\begin{equation}
\sum_{\mathsf P=\pi^0, \eta, \eta^\prime}  a_\mu^{{\rm HLbL}_{\mathsf P}}  \stackrel{{\rm uwa}} = 86.6(1.4) \times 10^{-11} .
\label{UWA}
\end{equation}
Plainly, our prediction has a significant impact on this average.
If one doubles the uncertainty on our result, the new average is:
\begin{equation}
\sum_{\mathsf P=\pi^0, \eta, \eta^\prime}  a_\mu^{{\rm HLbL}_{\mathsf P}}  \stackrel{{\rm iuwa}} = 88.9(1.5) \times 10^{-11} .
\label{IUWA}
\end{equation}

It is also worth remarking that if one drops the two earlier CSM studies, one obtains 
$87.8(1.1) \times 10^{-11}$.
Finally, for additional reference, the unweighted mean of the rather scattered plotted points is 
$89.7(4.4) \times 10^{-11}$.

In our view, given the sensitivity highlighted above and notwithstanding confidence in our approach and prediction, it is yet too soon to judge that\linebreak
$\sum_{\mathsf P=\pi^0, \eta, \eta^\prime}  a_\mu^{{\rm HLbL}_{\mathsf P}}$ is known with precision better than 5\%.  Further, collaborative analyses would be required to improve on that. 

Beyond the pseudoscalar meson pole HLbL scattering component $a_\mu$, there are other processes that may also provide a measurable contribution.  
For instance \cite[Sec.\,5]{Aliberti:2025beg}:
pseudoscalar meson ``box'' diagrams; meson-meson rescattering contributions; and pole contributions from states other than pseudoscalar mesons.
The methods utlised herein could also provide valuable, independent checks on existing estimates of such contributions, whether made using phenomenological tools, CSMs, or other nonperturbative approaches to QCD.

\begin{acknowledgements}
We are grateful for constructive comments from K.\ Raya.
Work supported by:
National Natural Science Foundation of China (grant nos.\ 12135007, 12505101).
\end{acknowledgements}

\appendix

\section{{\sf P}-meson pole HLbL contribution}
\label{AppendixA}
Working with the information in Sec.\,\ref{sectwo} and following Ref.\,\cite{Jegerlehner:2009ry}, one has the {\sf P}-meson pole HLbL contribution to the $\mu$-lepton anomalous magnetic moment:
\begin{align}
 a_\mu^{{\rm HLbL}_{\mathsf P}} & =
-\frac{2\pi}{3}\left(\dfrac{\alpha}{\pi}\right)^3
\int_0^\infty \! dk_1 dk_2 k_1^3 k_2^3
\int_{-1}^{1} du \sqrt{1-u^2} \nonumber \\
    & \quad \times \left[ \frac{F_{\mathsf P} (k_1^2,k_3^2) F_{\mathsf P}(k_2^2,0)}{k_2^2+m_{\mathsf P}^2} I_1(k_1, k_2, u) \right. \nonumber \\
& \qquad \left. + \frac{F_{\mathsf P}(k_1^2,k_2^2) F_{\mathsf P}(k_3^2,0)}{k_3^2 + m_{\mathsf P}^2} I_2(k_1,k_2, u)\right] .
    \label{Finalamu}
\end{align}

In Eq.\,\eqref{Finalamu},
$\alpha$ is the low-energy value of the fine structure constant in quantum electrodynamics;
$k_{1,2}$ are the magnitudes of the four-vectors in Eq.\,\eqref{Picoupling} and
$k_3^2 = k_1^2 + k_2^2 + 2 k_1 k_2 u$.
{\allowdisplaybreaks
Furthermore, with $Y_i = 1/k_i^2$:
\begin{align}
    I_1& = X(k_1,k_2,u) \nonumber \\
& \quad \times
    \bigg(8Y_1Y_2k_1k_2u - 2Y_1 Y_3\left[\frac{k_2^4}{m_\mu^2}-2k_2^2\right] \nonumber \\
    & \qquad -2Y_1\left[2-\frac{k_2^2}{m_\mu^2}+\frac{2k_1k_2u}{m_\mu^2}\right] +4Y_2Y_3k_1^2 \notag\\
    & \qquad - 4Y_2-2Y_3\left[4+\frac{k_1^2}{m_\mu^2}-2\frac{k_2^2}{m_\mu^2}\right] +\frac{2}{m_\mu^2}\bigg) \nonumber \\
&\quad  -2Y_1Y_2\left[1+(1-R_{m_1})\frac{k_1k_2 u}{m_\mu^2}\right]\notag\\
    & \quad  + Y_1Y_3 \left[2-(1-R_{m_1})\frac{k_2^2}{m_\mu^2}\right]+Y_1\frac{1-R_{m_1}}{m_\mu^2}\notag\\
    & \quad  +Y_2 Y_3 \left[2+(1-R_{m_1})^2\frac{k_1k_2 u}{m_\mu^2}\right] \nonumber \\
& \quad +3Y_3\frac{1-R_{m_1}}{m_\mu^2} , \\
%
    I_2 & = X(k_1,k_2,u) \nonumber \\
& \quad \times \bigg(4Y_1Y_2k_1k_2u + 2Y_1Y_3k_2^2 -2Y_1 \nonumber \\
& \qquad +2Y_2Y_3k_1^2 -2Y_2-4Y_3-\frac{4}{m_\mu^2}\bigg)\notag\\
    &\quad -2Y_1Y_2-3Y_1\frac{1-R_{m_2}}{2m_\mu^2}-3Y_2\frac{1-R_{m_1}}{2m_\mu^2} \nonumber \\
& \quad -Y_3\frac{2-R_{m_1}-R_{m_2}}{2m_\mu^2}\notag\\\
    & \quad +Y_1Y_3\left[2+3(1-R_{m_2})\frac{k_2^2}{2m_\mu^2} +(1-R_{m_2})^2\frac{k_1k_2 u}{2m_\mu^2}\right]\notag\\
    &\quad +Y_2Y_3 \left[2+3(1-R_{m_1})\frac{k_1^2}{2m_\mu^2}\right. \nonumber \\
    & \qquad \left. +(1-R_{m_1})^2\frac{k_1k_2 u}{2m_\mu^2}\right] , \\
    X & =\frac{1}{k_1 k_2\sqrt{1-u^2}} \arctan\left(\frac{z\sqrt{1-u^2}}{1-zu}\right) , \\
    z & =\frac{k_1k_2}{4m_\mu^2}(1-R_{m_1})(1-R_{m_2}) , \\
    R_{m_i} & =\sqrt{1+4\frac{m_\mu^2}{k_i^2}} .
\end{align}
Plainly, an $a_\mu^{HLbL_{\mathsf P}}$ prediction is available once the\linebreak $\gamma^{(\ast)}\gamma^{(\ast)} \mathsf P$ transition form factors, $F_{\mathsf P}=G_{\mathsf P}/[2\pi^2]$, are known, because all other factors are kinematic.  (We use experimental values for these numbers \cite[PDG]{ParticleDataGroup:2024cfk}.)
The transition form factor functions are detailed in Secs.\,\ref{SecThree}\,-\,\ref{Secetab}.
}

\begin{table*}[t]
\caption{ \label{currentmass}
Using the listed quark current-masses (Panel A\,-\,top row), one obtains the reported
RL truncation predictions for selected meson properties (Panel B).
As described in the text, regarding Eq.\,\eqref{defcalG}:
for light quark systems, $f=u, d, s$, one has $n_f=4$, $\Lambda_{\rm QCD} = 0.234\,$GeV, and $\omega = 0.5\,$GeV, $\varsigma_f  =0.8\,{\rm GeV}$; whereas in treating $c, b$ systems, $n_f=5$, $\Lambda_{\rm QCD} = 0.36\,$GeV, and $\omega = 0.8\,$GeV, $\varsigma_{c,b}  =0.6\,{\rm GeV}$.
Regarding Panel B, as explained in Sec.\,\ref{etaprimeeta}, $\eta$, $\eta^\prime$ mesons are characterised by two leptonic decay constants: $\{f^\ell,f^s\}$, each identifying the indicated quark flavour contribution.
The $\eta$, $\eta^\prime$ masses and decay constants are obtained with an anomaly kernel defined by 
$\surd D_\xi = 0.39\,$GeV,  
$\cos^2\theta_\xi = 0.94$, 
$r_A = 0.57$.
In Panel B, the rows labelled with a subscript ``emp'' provide comparisons with empirical values, inferred from Refs.\,\cite[PDG]{ParticleDataGroup:2024cfk}, \cite{Feldmann:1998sh, Benayoun:1999au, DeFazio:2000my}.
(All dimensioned quantities listed in GeV.)
}
\begin{tabular*}
{\hsize}
{
l@{\extracolsep{0ptplus1fil}}
|c@{\extracolsep{0ptplus1fil}}
c@{\extracolsep{0ptplus1fil}}
c@{\extracolsep{0ptplus1fil}}
c@{\extracolsep{0ptplus1fil}}}\hline\hline
\centering
A$\ $ & $m_\ell\ $ & $m_s\ $ & $m_c\ $ & $m_b\ $ \\\hline
$\hat m \ $ RGI $\ $ & $0.0067\ $ & $0.161\ $ & $1.61\ $ & $6.96\ $ \\
one-loop: $19\to 2\ $ & $0.0048\ $ & $0.112\ $ & $1.22\ $ & $5.26\ $ \\
$M_E$ (RGI) $\ $ & $0.406\phantom{8}\ $ & $0.566\ $ & $1.33\ $ & $4.12\ $ \\
\hline\hline
\end{tabular*}
\smallskip

\begin{tabular*}
{\hsize}
{
l@{\extracolsep{0ptplus1fil}}
|c@{\extracolsep{0ptplus1fil}}
c@{\extracolsep{0ptplus1fil}}
c@{\extracolsep{0ptplus1fil}}
c@{\extracolsep{0ptplus1fil}}
c@{\extracolsep{0ptplus1fil}}}\hline\hline
\centering
B$\ $ & $\pi^0 $ & $\eta\ $ & $\eta^\prime\ $ & $\eta_c\ $ & $\rule{-1ex}{0ex}\eta_b$ \\\hline
$m_{\mathsf P}\ $ & $0.134\ $ & $0.56\ $ & $0.96\ $ & $2.99\phantom{00}\ $ & $\rule{-1ex}{0ex}9.32\ $ \\
$\left. m_{\mathsf P}\right|_{\rm emp}\ $ & $0.135\ $ & $0.55\ $ & $0.96\ $ & $2.98\phantom{00}\ $ & $\rule{-1ex}{0ex}9.39\ $ \\
$f_{\mathsf P}\ $ & $0.093\ $ & $[0.073,-0.09]\ $ & $[0.069,0.105]\ $ & $0.28\phantom{00}\ $ & $\rule{-1ex}{0ex}0.56\ $ \\
$\left. f_{\mathsf P}\right|_{\rm emp}\ $   & $0.092\ $ & $[0.09(1),-0.09(3)]\ $  & $[0.07(1),0.09(1)]\ $  & $0.27(1)\ $ & \\
\hline\hline
\end{tabular*}
\end{table*}


\section{Rainbow-Ladder Kernel}
\label{AppendixRL}
Realistic implementations of RL truncation can be traced from Ref.\,\cite{Maris:1997tm}.
The principal element is the effective charge.
Studies of QCD's gauge sector have delivered the following form for the product of effective charge and gluon $2$-point function (propagator) \cite{Qin:2011dd, Binosi:2014aea, Binosi:2016wcx, Yao:2024ixu}:
\begin{align}
 {\mathpzc d}(y) & =
 \frac{2\pi}{\omega^4} D e^{-y/\omega^2} +\frac{\pi \gamma_m  \mathcal{F}(y)}{\tfrac{1}{2}\ln\big[ \tau+(1+y/\Lambda_{\rm QCD}^2)^2 \big]}\,,
 \label{defcalG}
\end{align}
where \cite{Qin:2018dqp, Yao:2021pyf}:
$\gamma_m = 4 /\beta_0 = 12/(33-2 n_f)$, with $n_f$ being the number of ultraviolet (UV) active quark flavours and
$\Lambda_{\rm QCD}$ is the associated RGI mass-scale;
and
$\tau={\rm e}^2-1$, ${\cal F}(y) = \{1 - \exp(-y/\Lambda_{\mathpzc I}^2)\}/y$, $\Lambda_{\mathpzc I}=1\,$GeV.

The kernel of the gap equation for any quark propagator, $S(q)$, can then be written as $(l=(q-q^\prime), y=l^2)$ \cite{Maris:1997tm, Binosi:2016wcx}:
\begin{equation}
\label{DSEkernel}
Z_2^2 {\mathpzc d}(y) T_{\mu\nu}(l) [i\gamma_\mu\frac{\lambda^{a}}{2} ]_{tr} [i\gamma_\nu\frac{\lambda^{a}}{2} ]_{su}\,,
\end{equation}
where $Z_2$ is the quark wave function renormalisation constant -- see below, and
$l^2 T_{\mu\nu}(l) = l^2 \delta_{\mu\nu} - l_\mu l_\nu$.
Landau gauge is used because, amongst other strengths, it is a fixed point of the QCD renormalisation group.  Once the kernel of the rainbow gap equation is specified, the RL Bethe-Salpeter kernel is determined by symmetry considerations \cite{Maris:1997tm}.

In solving all DSEs involved in the processes considered herein, we use a mass-independent momentum-subtraction renormalisation scheme \cite{Chang:2008ec}, with renormalisation scale $\zeta=\zeta_{2}:=2\,$GeV.  At $\zeta_{2}$, one finds that the flavour-independent quark wave function and current-mass renormalisation constants are
$Z_2 = 0.817$, $Z_4 = 0.281$.
Evolution to other renormalisation scales is straightforward.

Applications to many hadrons and processes have established \cite{Ding:2022ows} that interactions in the class typified by Eqs.\,\eqref{defcalG}, \eqref{DSEkernel} can unify the properties of a wide variety of systems.
Notably, when $\omega D=:\varsigma^3$ is held fixed, results for observables remain practically unchanged under $\omega \to (1\pm 0.1)\omega$ \cite{Qin:2020rad}.
Hence, the interaction in Eq.\,\eqref{defcalG} is determined by just one parameter: $\varsigma^3:=\omega D$.

In the light-quark sector, $f=u, d, s$, one has $n_f=4$, $\Lambda_{\rm QCD} = 0.234$, and finds that $\omega = 0.5\,$GeV, $\varsigma_f  =0.8\,{\rm GeV}$ deliver a good description of a range of pseudoscalar and vector meson static properties. 
For heavier quarks, $n_f=5$, $\Lambda_{\rm QCD} = 0.36$, and, as noted above, it is known that corrections to RL truncation are less important -- see, \emph{e.g}.,  Ref.\,\cite{Bhagwat:2004hn} for illustration, so the interaction parameter should be closer to that used in more sophisticated kernels, \emph{viz}.\ $\omega = 0.8\,$GeV, $\varsigma_{c,b}  =0.6\,{\rm GeV}$ \cite{Qin:2018dqp, Yao:2021pyf}.
This is discussed further in, \emph{e.g}.,\linebreak Ref.\,\cite[Sect.\,IIB]{Xu:2019ilh}.

With the RL interaction thus defined, the quark current masses may be determined by requiring a good description of a few meson properties, such as pion and kaon masses and decay constants, and $\eta_c$, $B_c$ masses.  Following Refs.\,\cite{Ding:2018xwy, Yao:2021pyf}, this yields the values in Table~\ref{currentmass}\,A.
They provide a fair match with values obtained via other means \cite[PDG]{ParticleDataGroup:2024cfk}

\section{Triangle Diagram Kinematics}
In evaluating the integral in Eq.\,\eqref{anomalytriangle} or \eqref{etaIntegral}, we use two different choices for the external momenta when calculating $F_{\mathsf P}(t,0)$.  The frame defines the momenta used when solving the associated meson Bethe-Salpeter equations.  

\subsection{Moving Frame}
\label{AppendixMoving}
{\allowdisplaybreaks
For lighter mesons, the ``moving frame'' maximises the domain upon which direct calculation of $F_{\mathsf P}(t,0)$ is possible.   Namely, with $t_{12}=t_1+t_2$, $\check t_{12}=t_1-t_2$, 
\begin{subequations}
\begin{align}
k_1 & = (0,0,0,\surd t_1) , \\
P & = -k_1-k_2 = (0,0,P_3,P_4) , \\
P_3 & = -i \frac{m_{\mathsf P}^2}{2 \surd t_1} \sqrt{1+2 t_{12}/m_{\mathsf P}^2 + \check t_{12}^2/m_{\mathsf P}^4} ,\\
P_4 & = \frac{m_{\mathsf P}^2 -\check t_{12}}{2 \surd t_1} .
\end{align}
\end{subequations}
This is the choice appropriate for $\pi^0$, $\eta$.  It enables one to compute directly on the following domains:
\{$\pi^0  \, | \, 0.04 \leq t/{\rm GeV}^2 \leq 3.0$\};
\{$\eta \, | \, 0.04 \leq t/{\rm GeV}^2 \leq 2.5$\}.
}

\subsection{Rest Frame}
\label{AppendixRest}
{\allowdisplaybreaks
For the heavier mesons, \emph{viz}.\  $\eta^\prime$, $\eta_{c}$, $\eta_b$, the meson rest frame is best:
\begin{subequations}
\begin{align}
P & = k_1 + k_2 = (0,0,0, i m_{\mathsf P}) , \\
k_1 & =  (0,0,k_{13},k_{14}) ,\\
k_{13} & = \tfrac{1}{2}\sqrt{\check t_{12}^2/m_{\mathsf P}^2 + m_{\mathsf P}^2 + 2 t_{12}} ,\\
k_{14} & = - i \frac{m_{\mathsf P}^2 - \check t_{12}}{2 m_{\mathsf P}} . 
\end{align}
\end{subequations}
This frame provides direct calculation access to the following domains:
\{$\eta^\prime \, | \, 0.0 \leq t/{\rm GeV}^2 \leq 1.0$\};
\{$\eta_{c} \, | \, 0.0 \leq t/{\rm GeV}^2 \leq 9.0$\};
\{$\eta_{b} \, | \, 0.0 \leq t/{\rm GeV}^2 \leq 81.0$\}.
}

\subsection{Comments}
Regarding $\eta^\prime$, the moving frame enables direct calculation access on \{$\eta^\prime \, | \, 0.5 \leq t/{\rm GeV}^2 \leq 1.7$\}.  This measures less than the domain available to rest frame calculations; nevertheless, on the overlap domain, the results are identical. 

Owing to the balanced kinematics, $F_{\mathsf P}(t,t)$ can be calculated in either frame.  The ``rest frame''  is used because it is simplest.


\end{document}